\documentclass[twocolumn,showpacs,amssymb,prb]{revtex4}
\newcommand{\be}{\begin{equation}}
\newcommand{\ee}{\end{equation}}
\newcommand{\tw}{t_{\rm w}}

\newcommand{\ximin}{\xi_{\rm min}}
\newcommand{\teff}{t_{\rm eff}}
\newcommand{\xicool}{\xi_{\rm cool}}
\newcommand{\tcool}{t_{\rm cool}}
\newcommand{\Tcool}{T_{\rm cool}}
\newcommand{\teffcum}{t_{\rm eff}^{\rm cum}}
\usepackage{graphicx}
\usepackage{epsfig}

\begin{document}

\title{Temperature Cycles in the Heisenberg Spin Glass}

\author{L.~Berthier}
\email{berthier@lcvn.univ-montp2.fr}
\homepage{http://w3.lcvn.univ-montp2.fr/~berthier}
\affiliation{Laboratoire des Collo\"{\i}des, Verres 
et Nanomat\'eriaux, 
Universit\'e Montpellier II and UMR 5587 CNRS, 34095
Montpellier Cedex 5, France}

\author{A.~P.~Young}
\email{peter@bartok.ucsc.edu}
\homepage{http://bartok.ucsc.edu/peter}
\affiliation{Department of Physics,
University of California,
Santa Cruz, California 95064, USA}

\date{\today}

\begin{abstract}
We study numerically the nonequilibrium dynamics of the three-dimensional
Heisenberg Edwards-Anderson spin glass submitted to 
protocols during which temperature is shifted or cycled within 
the spin glass phase. We show that (partial) rejuvenation and 
(perfect) memory effects can be numerically observed and 
study both effects in detail. We quantitatively characterize 
their dependences on parameters such as 
the amplitude of the temperature changes, the timescale at which the 
changes are performed, and the cooling rates used to vary the temperature.
We contrast our results both to those found 
numerically in the Ising version of the model,
and to experimental results in different samples.
We discuss the theoretical interpretations of our findings, arguing, in
particular, that `full' rejuvenation can be observed in
experiments even if temperature chaos is absent.
\end{abstract}

\maketitle

\section{Introduction to basic phenomena}
\label{introduction}

\begin{figure}[b]
\begin{center}
\psfig{file=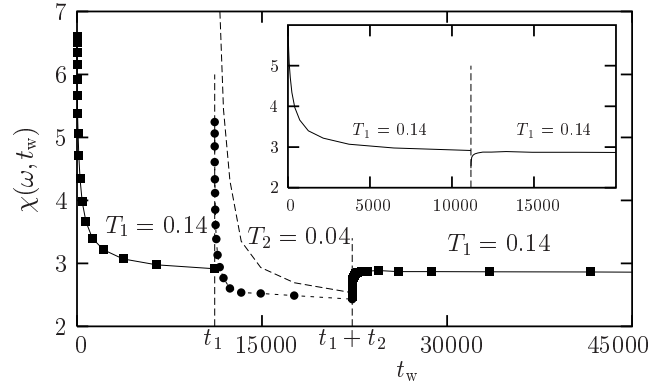,width=8.5cm}
\caption{\label{cycle} 
Time evolution of $\chi(\omega, \tw)$ in a temperature cycle 
performed with the three-dimensional Heisenberg spin glass for 
$1/\omega$ = 1078.
Aging is observed in the first $t_1 = 11159$ 
sweeps during which the temperature is 
at $T_1=0.14 < T_c \simeq 0.16$. The temperature is
then quenched to $T_2 = 0.04$
for the next $t_2 = 11159$ sweeps.
This causes aging to restart, a rejuvenation effect.
The long-dashed line shows the result of a
direct quench from $T=\infty$ to $T_2$ at time $t_1$.
The variation of 
$\chi(\omega,\tw)$ with $\tw$
is much stronger than when the temperature is shifted from $T_1$ to $T_2$,
showing that rejuvenation in the temperature shift
is only partial. At $\tw=t_1+t_2$ the temperature is shifted back to 
its initial value, and the data seems to be simply a continuation
of that for $\tw<t_1$. This
memory effect is illustrated more clearly in the inset where
the intermediate time spent at $T_2 = 0.04$ has been removed.
}
\end{center}
\end{figure}

In recent work~\cite{I} we studied the nonequilibrium dynamics
of the Heisenberg Edwards-Anderson spin glass model in three dimensions
following a sudden quench to its low temperature phase.
Here we continue our investigations of the nonequilibrium, 
low temperature dynamics of this model
by analyzing its behavior 
in more complex protocols, similar to those 
performed in experiments~\cite{review1,review1_1,review3}.
More precisely, we consider in detail the effects of a temperature cycle
(illustrated in Fig.~\ref{cycle})
consisting of the following three steps:
\begin{enumerate}
\item
Quench the system from $T=\infty$ to a temperature $T_1$ at time $\tw = 0$,
and wait a time $t_1$.
\item
Then
change the temperature to a lower value $T_2$ and wait a further time
$t_2$. 
\item
At total time $\tw = t_1 + t_2$
change the temperature back to $T_1$.
\end{enumerate}
This temperature cycling protocol has been used extensively to characterize
several spin glass 
systems~\cite{review1,review1_1,review3,dupuis,yosh4,yosh3,bert,cycle,physica,dip,petra,kovacs,india,matthieu},
and these studies 
motivated similar cycling experiments on many different types of glassy 
materials~\cite{struik,frustre,leheny1,leheny2,beta,KLT,bellon,ferro,ferro2,electron}. 

Numerical studies of temperature shifts and cycles in 
the Ising version of the 
model have
been reported~\cite{heiko,rieger,yosh2,BB1,BB2,ricci,ricci2,jimenez}, 
and temperature cycles have been discussed 
theoretically~\cite{fh,jorge,jp,encorejp,yosh,surf,yosh5_1,martin,yosh5}.
However, the Heisenberg spin glass model has been much less studied than its
Ising counterpart, even though
it is closer to many of the experimentally studied spin glass systems,
and only a few 
aging studies exist~\cite{I,II,ricci2,kawa}. 
In particular, extensive studies of temperature cycles have not been
reported (see Ref.~[\onlinecite{ricci2}] for preliminary results)
and we attempt here to fill this gap.

The Heisenberg spin glass model that we study has 
the Hamiltonian 
\begin{equation}
H = - \sum_{\langle i,j \rangle} J_{ij} {\bf S}_i \cdot {\bf S}_j,
\label{ham}
\end{equation}
where the sum is over nearest neighbors of a cubic lattice
of linear size $L=60$ with periodic boundary conditions,
the ${\bf S}_i$ are three-component vectors of unit length and 
coupling constants are drawn from a symmetric 
Gaussian distribution of standard deviation unity.
We refer to our previous paper~\cite{I} 
for the technical details concerning our simulations.
When performing
temperature cycles a large number of parameters can be varied,
implying a larger numerical effort than was needed in the simple aging
studies of Ref.~[\onlinecite{I}].

The basic quantity we measure is the two-time autocorrelation function of the
spins,
$C(\tw, \tw + \tau)$, defined by
\begin{equation}
C(\tw, \tw + \tau) = \frac{1}{L^3} \sum_i \langle  {\bf S}_i(\tw) \cdot
{\bf S}_i(\tw +\tau) \rangle,
\label{auto}
\end{equation}
where the brackets indicate an average over both thermal histories and
disorder. To compare with experiment we often plot instead
\begin{equation}
\chi(\omega,\tw) \equiv \frac{1- C(\tw, \tw + \omega^{-1})}{T},
\label{sus}
\end{equation}
since this is expected to have similar behavior to the ac 
magnetic susceptibility at frequency $\omega$.

A representative sample of our results for a temperature cycle is shown in
Fig.~\ref{cycle}. For $\tw < t_1$ this is just a `simple aging' experiment
corresponding to quenching the system infinitely fast from an infinite to a
low temperature, $T_1 = 0.14 < T_c \simeq 0.16$.  Physical quantities then
slowly relax towards equilibrium, which is known as aging.  We have
characterized  this behavior in detail in Ref.~[\onlinecite{I}].

For $t_1 < \tw < t_1 + t_2$, during which
the temperature is $T_2 = 0.04 < T_1$,
the signal just after the shift is not a
simple continuation of the decay at the previous temperature. 
The system has apparently forgotten it is already `old' and it seems
therefore `rejuvenated' by the temperature change. 
In experiments performed on Heisenberg spin glasses,
when $T_1 - T_2$ is sufficiently large,
the signal obtained after the shift can be exactly superposed 
on the one obtained after a direct quench from $T=\infty$ to $T=T_2$,
implying that the system behaves as if it had fully forgotten 
the time spent at temperature $T_1$.
We will call this situation
`full rejuvenation'.
However,
rejuvenation is not full in our simulations
because the change in $\chi(\omega, \tw)$ with
time during the part of the temperature
cycle at $T_2$ is less than the change found in a
direct quench to $T_2$, which is 
shown by the long-dashed line in Fig.~\ref{cycle}. 
Since we do not observe full rejuvenation we will need to ask
whether the observed
behavior is rejuvenation at all or whether it can be fully
explained in terms of the cumulative aging scenario, discussed below
in Sec.~\ref{cumag}. We will also investigate the 
possible origins of this significant difference between 
numerical and experimental results, and will 
discuss in particular the role of finite cooling rates 
in experiments in Secs.~\ref{direct} and \ref{disc:fullrejuv}.

Finally, in the third part 
of the experiment, $\tw > t_1 + t_2$,
the relaxation proceeds, after a short transient,
as if the second step had not occurred. The system has kept 
a `memory' of the first aging step despite the strong restart observed
in the second part of the protocol.
In the inset of Fig.~\ref{cycle}, we show that the third 
part of the experiment appears to be the simple 
continuation of the first part, as if 
the second step had not taken place, which we 
call a `perfect memory effect'. This figure is not entirely 
convincing, though, 
because the signal is nearly flat and small deviations 
from perfect memory could be invisible on that scale. 
That memory is indeed close to perfect, at least for large temperature shifts,
will be more quantitatively 
demonstrated in Sec.~\ref{memory}.

In the rest of this work, we analyze in more detail 
the rejuvenation and memory effects seen in Fig.~\ref{cycle}.
Section \ref{cumag} gives some theoretical background on
the cumulative aging scenario,
according to which the effects seen are simply due to changes in the rate
of growth of correlations with temperature. Our data on
rejuvenation and memory are 
described in Secs.~\ref{rejuvenation} and~\ref{memory} respectively.
We discuss our results in Sec.~\ref{discussion}, and give a final summary in
Sec.~\ref{summary}. 

\section{Cumulative Aging}
\label{cumag}

An important theoretical concept in understanding non-equilibrium data for
spin glasses is that of 
a dynamical correlation 
length, $\xi(T,t)$, which describes the spatial extent of spin correlations
after aging a time
$t$ at temperature $T$. In the cumulative aging scenario, it is proposed
that the observed phenomena can be understood solely
in terms of the time and
temperature dependences of $\xi(T, t)$. We now discuss the predictions of
cumulative aging, first
for the step down in temperature from $T_1$ to $T_2$ which
can lead to rejuvenation, and then for the subsequent step back up to $T_1$
which can lead to memory.

\subsection{Cumulative aging and rejuvenation}
\label{cumagt1}

If the correlation length grows to a value $\xi(T_1,t_1)$ by waiting a time
$t_1$ at $T_1$, then one would have to wait a longer time, $\teffcum$,
at the lower temperature $T_2$ to get the same correlation length, 
where this `effective
age' in the cumulative aging scenario is given by
\begin{equation}
\xi(T_1,t_1) = \xi(T_2,\teffcum).
\label{teffeq}
\end{equation}
If the growth law
$\xi(T,t)$ is known accurately, Eq.~(\ref{teffeq}) can be used to determine
$\teffcum$ for given values of $T_1, T_2$ and $t_1$.

In practice one can try to fit 
the data for $\chi(\omega, t_1 + t)$ during the
time spent at $T_2$ to data for
$\chi_{_{T_2}}(\omega, \teff + t)$,
i.e.~data in a single quench to $T_2$ for some
waiting time $\teff + t$ with $\teff$ used as a free fitting
parameter. In other words
$\teff$ is determined from
matching both sides of the following equation, 
\begin{equation}
\chi(\omega, t_1 + t) = \chi_{_{T_2}}(\omega, \teff + t) \, .
\label{chicumag}
\end{equation}

One of the following scenarios might then occur. A first possibility
is that
\begin{equation}
\teff = 0  \, ,
\end{equation}
which corresponds to full rejuvenation.
The time spent at $T_1$ is irrelevant for aging at
$T_2$. This is seen experimentally for large temperature shifts,
but not in any of our simulations.

Second, one might find that
\begin{equation}
\teff = \teffcum \, ,
\label{tprimetf}
\end{equation}
where $\teffcum$ is given by Eq.~(\ref{teffeq}). This is 
the so-called cumulative aging scenario. 
We shall see in Sec.~\ref{memory_heis} that
this case describes our
simulations when the temperature difference $T_1 - T_2$ is small, though
we show in Sec.~\ref{rej_data} that it does not work when $T_1-T_2$
is large. 
However, Eq.~(\ref{tprimetf}) does
not correspond to
rejuvenation. Rather, rejuvenation is defined to be the additional
effect beyond Eq.~(\ref{tprimetf}), since the latter simply reflects
the change in the rate of growth of the correlation
length with temperature.

A third possibility is that 
\begin{equation}
0 < \teff < \teffcum \, .
\end{equation}
This is one way in which partial rejuvenation could occur, and is seen
in some experiments~\cite{yosh4,yosh3,petra} at intermediate values of the
temperature shift.
However our simulations do not fit this scenario.
Rather, for large values of $T_1 - T_2$ we will see in 
Sec.~\ref{rej_data} that
our data cannot be collapsed on to data for a direct quench by shifting the
time. Hence the effective age $\teff$ cannot be defined 
in our simulations at large temperature differences, 
and so our results in this region do not fit any of the above scenarios.

Remark that if $t_1$ and $T_1-T_2$ are
sufficiently large, $\teffcum$ will be much larger than $t_1$.
One can make this statement
more quantitative by assuming, as usual, that equilibration proceeds
by activation over barriers, 
where the barrier height $\Delta E$
is some function of the length scale $\xi$. 
Writing $t = t_0 \exp( \Delta E(\xi) / T)$, where $t_0$ is a microscopic 
attempt time which we set to 1, Eq.~(\ref{teffeq}) gives
\begin{equation}
T_1 \ln t_1 = T_2 \ln \teffcum \, ,
\label{T1lnt1}
\end{equation}
which can be written as
\begin{equation}
\ln \left(
{\teffcum \over t_1} \right)
= \left({T_1 - T_2 \over T_2}\right) \ln
t_1  \, .
\label{teffcumt1}
\end{equation}
Hence if $(T_1 - T_2) \ln t_1 > T_2$ we have $\teffcum \gg t_1$. As a
result, $\chi(\omega, t_1 + t)$, given by Eq.~(\ref{chicumag}),
will vary very little with $t$ unless $t$ greatly exceeds $t_1$, which is not
the case in the simulations. 
Although the reasoning leading to 
Eqs.~(\ref{T1lnt1}) and (\ref{teffcumt1}) is for
a particular simple model of the growth of the correlation
length, we expect the result $\teffcum \gg t_1$ to be 
more generally correct.

\subsection{Cumulative aging and memory}
\label{cumagt2}

In a similar way we can fit the data for $\chi(\omega, t_1 + t_2 + t)$ after
the step back up to $T_1$ to data for
$\chi_{_{T_1}}(\omega, t_1 + \teff + t)$,
\begin{equation}
\chi(t_1 + t_2 + t) = \chi_{_{T_1}}(\omega, t_1 + \teff + t) ,
\label{chimemstepup}
\end{equation}
assuming that the two functions of $t$ in this expression 
can be matched for a particular choice of $\teff$.
If memory is perfect then
\begin{equation}
\teff = 0 \, ,
\end{equation}
so that the time spent at $T_2$ simply plays no role 
in the subsequent aging at $T_1$.

In the cumulative aging scenario $ \teff = \teffcum $ where
\begin{equation}
\xi(T_1, t_1 + \teffcum ) = \xi(T_2, t' + t_2) \, ,
\label{xi1}
\end{equation}
in which $t'$ is the time the system would have to age at $T_2$ to get the
correlation length it reached at $T_1$ after time $t_1$. It is 
given implicitly by
\begin{equation}
\xi(T_2, t') = \xi(T_1, t_1) \, ,
\label{xi'}
\end{equation}
as in Eq.~(\ref{teffeq}) above. This is shown graphically for a set of
experimental parameters in Fig.~\ref{th} below.

It is worth mentioning that at extremely long times or
for very large changes in temperature one expects perfect 
memory even in the cumulative
aging scenario. This is because, at long times, growth of $\xi$ with time
at the lower
temperature $T_2$ is much slower than at $T_1$, so $t'$ is huge, the 
difference
$\xi(T_2, t' + t_2) - \xi(T_2, t')$ is small, and hence $\teffcum$ is small.
As in Sec.~\ref{cumagt1}, we can make this more quantitative within a barrier
activation model for which
Eqs.~(\ref{xi1}) and (\ref{xi'}) imply
\begin{eqnarray}
T_2 \ln (t' + t_2) & = & T_1 \ln (t_1 + \teffcum)
\label{T2lnt2} \, , \\
T_2 \ln t' & = & T_1 \ln t_1 \,
\label{T2lnt'} .
\end{eqnarray}
We expect perfect memory if 
\begin{equation}
t' \gg t_2 \qquad \mbox{(perfect\ memory).}
\label{t'ggt2}
\end{equation}
Since $t_1 \sim t_2$,
from Eq.~(\ref{T2lnt'}) this gives
\begin{equation}
{T_1 - T_2 \over T_2} \gg {1 \over \ln t_1} \qquad \mbox{(perfect\ memory)} .
\label{memperf}
\end{equation}
Eq.~(\ref{t'ggt2}) is equivalent to the condition $\teffcum \ll t_2$ since
eliminating
of $t'$ from Eqs.~(\ref{T2lnt2}) and (\ref{T2lnt'}),
gives
\begin{equation}
{\teffcum \over t_2} = {T_2 \over T_1} \exp\left[-\left({T_1 - T_2 \over
T_2}\right) \ln t_1\right]  \, ,
\end{equation}
assuming $\teffcum \ll t_1$. Hence, if Eq.~(\ref{memperf}) is satisfied, one has
perfect memory in the cumulative aging scenario.  This conclusion should again
be more general than the particular barrier activation model we used.

\section{Rejuvenation}
\label{rejuvenation}

Now we discuss in detail our results for the behavior following
the temperature drop to $T_2$ and their interpretation.

\subsection{Do we really observe rejuvenation?}
\label{rej_data}

\begin{figure}
\begin{center}
\psfig{file=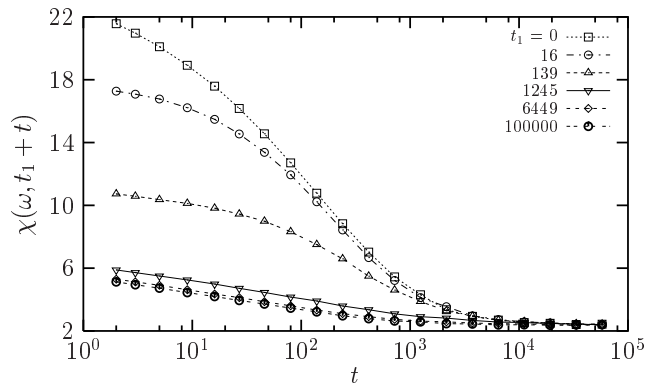,width=8.5cm}
\caption{\label{rejuv} 
The behavior of $\chi(\omega, t_1+ t)$ during the time $t$ spent at 
temperature $T_2 = 0.04$
for different values of the time, $t_1$, waited at $T_1 = 0.14$.
We fix $1/\omega = 1078$.  For $t_1=0$, the system undergoes a direct
quench to $T=T_2$, while for $t_1\to\infty$, the system has time to
equilibrate at $T_1$.
The restart of aging (rejuvenation)
is reduced when $t_1$ increases from 0, but 
remains non-zero even when $t_1 \to \infty$.}
\end{center}
\end{figure}

It is clear from Fig.~\ref{cycle} that aging 
is strongly restarted when temperature changes from
$T_1=0.14$ to $T_2=0.04$, but that, unlike in experiments,
the effect is weaker than when the system
is directly quenched to $T_2$. 
In Fig.~\ref{rejuv} we show the systematic trend in the data when we vary the
time $t_1$ spent at the upper temperature $T_1$.
Two important pieces of information can be deduced from Fig.~\ref{rejuv}. 

\begin{itemize}

\item
The case of $t_1=0$ corresponds to a direct quench from
$T=\infty $ to $T=T_2$.
We see from Fig.~\ref{rejuv}
that the signal obtained after waiting for a finite value of $t_1$ at
$T=T_1$ is
different from the signal obtained in a direct quench and differs more as
$t_1$ is increased. In `full rejuvenation', as seen in experiments, the data
after waiting a time $t_1$ is the same as in a direct quench.
We were not able to vary parameters of temperature 
cycles in such a way that full rejuvenation is observed in our simulations.

\item
When $t_1$ increases, the signal saturates
to a limiting behavior. This is
actually the region in which
the intermediate part of Fig.~\ref{cycle} was
obtained, since $t_1$ is quite large
there.
This saturation
implies that even if we were able to equilibrate 
the system at temperature $T_1$, it would undergo 
aging when further quenched to $T_2$. This is very different
from the behavior of a pure ferromagnet, for instance,
which would reequilibrate on a microscopic timescale
upon a similar temperature change within its ferromagnetic phase.  

\end{itemize}

\begin{figure}
\begin{center}
\psfig{file=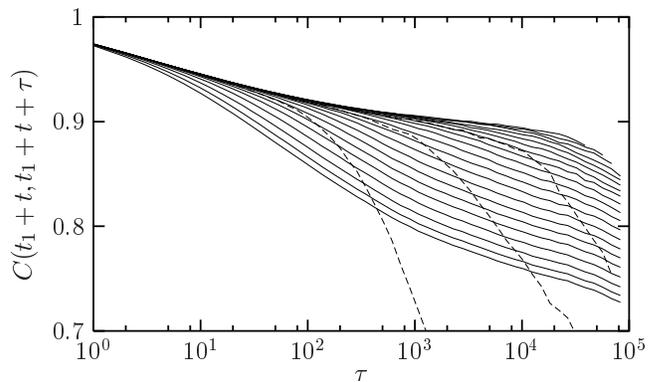,width=8.5cm}
\caption{\label{reju}
The full lines show data for the two-time spin autocorrelation
function $C(t_1 + t, t_1 + t + \tau)$ as a function of the time difference
$\tau$.
The time spent at $T_1=0.14$
is $t_1=10^5$ and the different curves
are for different values of $t$, the time waited at $T_2=0.04$, increasing from
$t=2$ (the lowest) to $t=57797$ in a logarithmic progression. This data shows
that aging is restarted in a temperature shift.
The dashed lines are the autocorrelation function 
measured in a direct quench to
$T_2=0.04$ and waiting times $t=416$, 3728 and 19307 
(from left to right). A comparison shows
that the shape of the
data found in reducing the temperature from $T_1$ to $T_2$
is different from that in the direct quench to $T_2$.}
\end{center}
\end{figure}

We have therefore seen that
rejuvenation is not full. We will now argue that it is not null either
and that there is an additional signal beyond that expected in the cumulative
aging scenario discussed in Sec.~\ref{cumag}.
To do so it is useful to
look at the dependence of $\chi(\omega, t_1+t)$
on frequency, or
equivalently the dependence of
the two-time autocorrelation function on the time difference $\tau$.
The solid lines in Fig.~\ref{reju} show data for this quantity.
The system has spent a large time
$t_1=10^5$ at $T=T_1=0.14$, sufficiently large that
the data in Fig.~\ref{reju}
has become independent of $t_1$.
The different curves
are for different values of $t$, the time waited at $T_2=0.04$, increasing from
$t=2$ to $57797$ in a logarithmic progression. 
The curves do not superpose 
and the behavior observed in Fig.~\ref{reju} is 
qualitatively similar to that obtained in simple aging experiments
where samples undergo a rapid quench to the spin glass phase. There is
a first, fast stationary decay of the autocorrelation functions followed by a
second, much slower, age-dependent decay.
However, there is a quantitative difference since the
precise shape of the data is actually not the same as in a direct
quench to $T_2$. The difference can
be appreciated in Fig.~\ref{reju} by comparing the solid with the
dashed lines, which are for a direct quench with various waiting times. In the
direct quench, the correlation function decays faster at long time
differences.

For the temperature shift to $T_2$
in Fig.~\ref{cycle}, we have seen in Fig.~\ref{reju} that the shape
of the extra signal due to this shift is different from the result of
a direct quench to $T_2$. We also need to ask whether the magnitude
of the extra signal is greater than that expected in the cumulative aging
scenario discussed in Sec.~\ref{cumag}. 

In Fig.~\ref{cycle}, the time 
$t_1 = 11159$ is
large, since the data in Fig.~\ref{rejuv} has saturated for this value
of $t_1$, and the
temperature shift $T_1 - T_2$ is large. Hence, as discussed in
Sec.~\ref{cumagt1}, the effective age in the cumulative aging scenario is
enormous, and the time dependence of $\chi$ 
should be tiny in this scenario. 
This is clearly not the case for the data in Fig.~\ref{cycle}.
We conclude that for large temperature shifts there is genuine rejuvenation; 
the cumulative aging scenario does not work.
In addition, the fact that the data of
Fig.~\ref{reju} is independent of $t_1$, at large $t_1$, for all $\tau
\equiv\omega^{-1}$,
implies that the above conclusion on the existence 
of rejuvenation holds for any frequency
$\omega$ accessible in our simulations. 

\begin{figure}
\begin{center}
\psfig{file=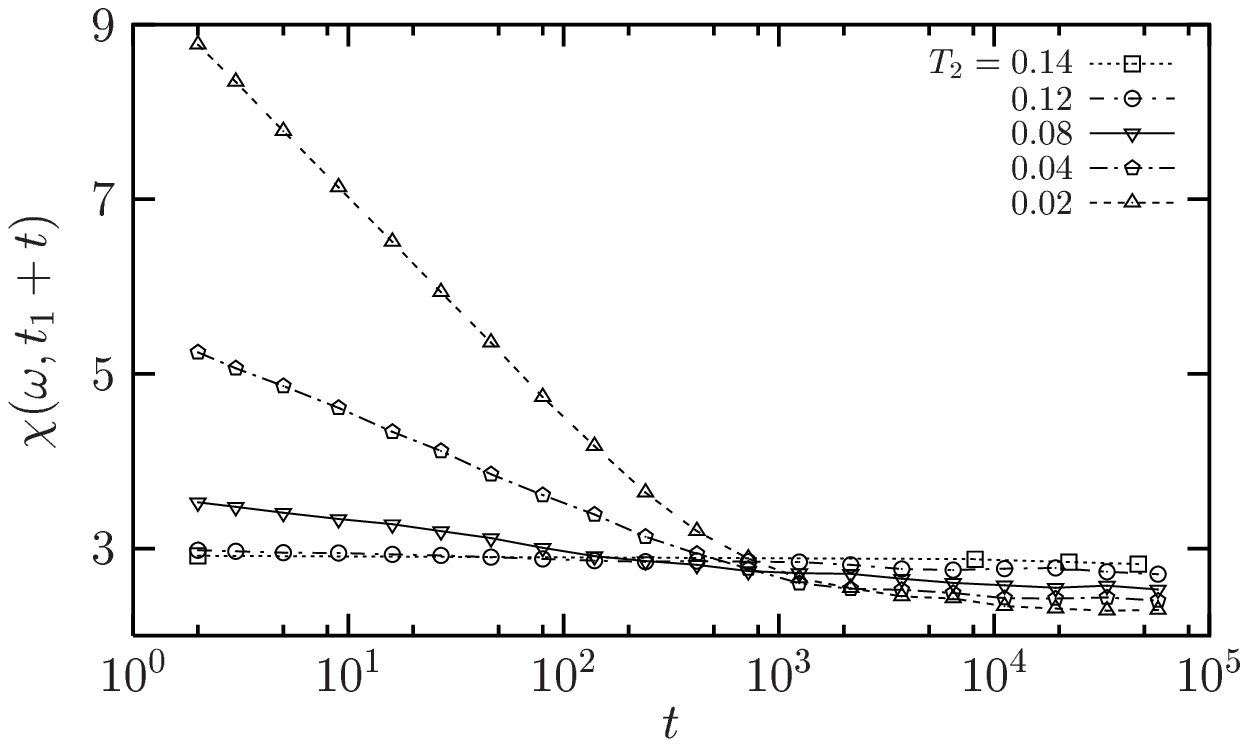,width=8.5cm}
\psfig{file=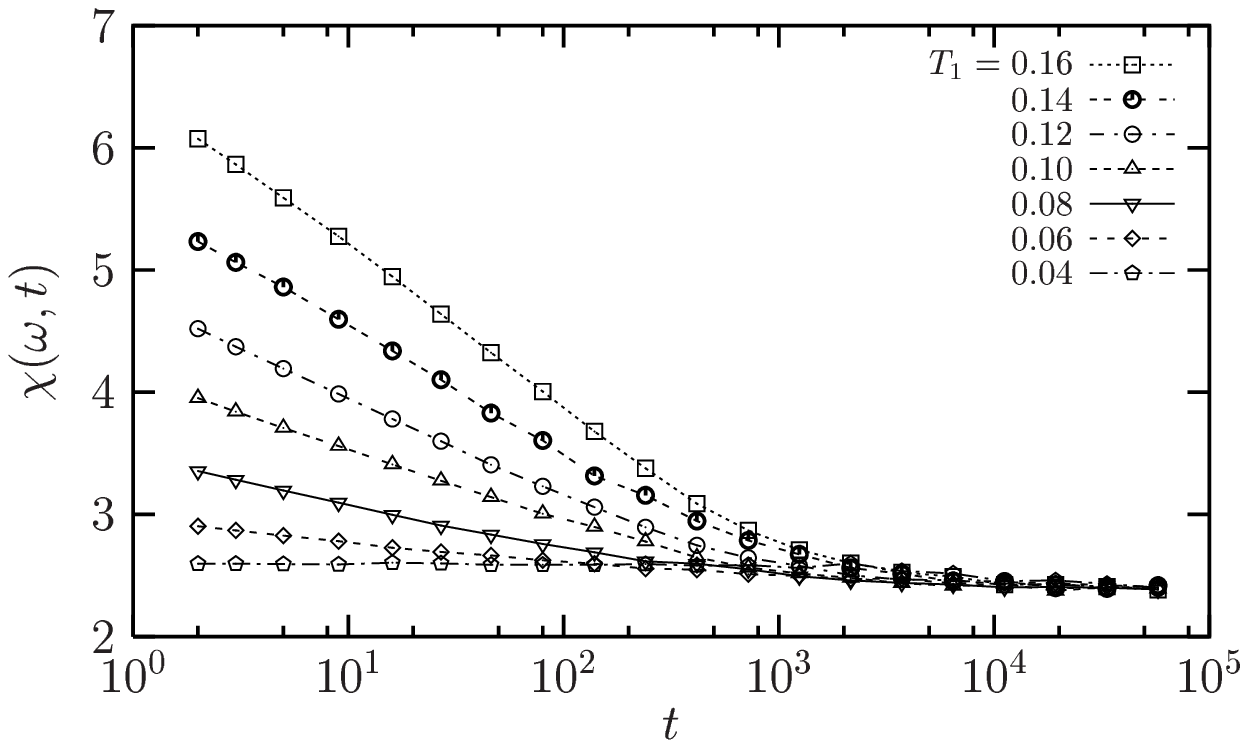,width=8.5cm}
\caption{\label{rej}
Top: Changing the second temperature.
The protocol is $T=\infty \to T_1=0.14$ at time 0. Temperature
is kept constant for $t_1=11159$ sweeps after which it is shifted 
to various $T_2 \le 0.14$.
Bottom: Changing the first temperature. The protocol and notations are as
above but with a fixed $T_2 = 0.04$ and different $T_1 \ge T_2$
In both figures the additional amplitude of the aging signal
increases with $T_1 - T_2$. 
While cumulative aging works for small values of $T_1 - T_2$, 
see Sec.~\ref{memory}, there is a genuine additional 
rejuvenation signal at large temperature differences.} 
\end{center}
\end{figure}

We note that the use 
of a different `clock' to determine effective ages 
could lead to the wrong impression that we observe 
full (or close to full)
rejuvenation~\cite{yosh3,BB2,dlogS}.
If one follows a common experimental procedure and defines
$\teff$ as the location of a peak in 
the logarithmic derivative of the two-time autocorrelation
function, $S(t, t+\tau) = \partial C(t, t +\tau) / \partial \log \tau$, 
one obtains a very small effective age from the data of
Fig.~\ref{reju} and then conclude that 
rejuvenation is almost full. We know, however, for
example by comparing the shapes of the
solid and dashed curves in Fig.~\ref{reju}, that this is not the case.
Hence $S(t, t+\tau)$ can not be used 
to quantify rejuvenation in numerical simulations
of temperature shifts.
This remark was already made for the Ising spin glass
in Ref.~[\onlinecite{BB2}], although  
in a less detailed way.

Data for a range of temperature shifts is presented in Fig.~\ref{rej}. We see
that the amplitude of the restart of aging is strongly dependent on the
amplitude of the temperature shift. The case where $T_2 = T_1$ has no restart
of aging and corresponds to
a simple quench to $T_2$ (with the plot starting at waiting
time $t_1$). For $T_2 < T_1$ there is an additional decay of $\chi(\omega, t)$
due to the temperature shift.
We shall see in Sec.~\ref{memory_heis}
that this effect is describable by
cumulative aging for small temperature shifts, unlike the situation for large
temperature shifts discussed earlier in this section.

As a final comment, we note that
the restart of aging observed in the second part of the cycle shown 
in Fig.~\ref{cycle} is not an artifact produced by our 
choice of $\chi(\omega,t)$ as a physical observable, 
as suggested in Refs.~[\onlinecite{ricci2,jimenez}]. It is certainly true that 
the susceptibility, defined by Eq.~(\ref{sus}) and the
correlation function
are not related by a fluctuation-dissipation relation
since the system is not in equilibrium, and so
rejuvenation effects might appear to be stronger using one or the 
other observable.
However, Fig.~\ref{reju} leaves us with no doubt concerning
the fact that aging is restarted in a shift. 

\subsection{Is temperature chaos relevant?}
\label{secchaos}

In this subsection we discuss our results about rejuvenation effects
from a theoretical point of view.
As already mentioned, the growth of correlations with time is characterized by
a dynamical correlation length $\xi(T, t)$.
Lengthscales that are active at
$T_1$ in a given time window $t$, $\xi_1 \equiv \xi(T_1,t)$,
are typically
larger than the ones active at $T_2<T_1$ in the same time window,
$\xi_2 \equiv \xi(T_2,t) < \xi_1$.
Although $\xi(T,t)$ is only a gradual function of $T$ at fixed time $t$,
$\xi$ grows very slowly (presumably logarithmically)
with $t$ and so
it can take an astronomical amount of time
to relax excitations of size $\xi_1$
at the lower temperature $T_2$.
Roughly speaking,
changing the temperature
gives only a modest change in length scales but a huge change in time scales. 
For practical purposes, excitations of size $\xi_1$ are frozen at $T_2$
and dynamics at $T_2$ is due to
fluctuations at the smaller length scale $\xi_2$,
which had easily
reached quasi-equilibrium at temperature $T_1$. 
Rejuvenation therefore means that
equilibrium states at $T_1$ and $T_2$
are significantly `different'
on length scales of the order of $\xi_2$.
What `different' really means will be the subject of this
subsection.

Theoretically two explicit examples of this difference have been discussed
in the literature:
\begin{enumerate}
\item
Temperature chaos in the scaling picture of 
spin glasses~\cite{fh,bm1,timo,hajime,kr}.
\item
Temperature dependence of the exponent of the
power-law decay of spatial correlations in the two-dimensional
non-disordered XY model~\cite{surf}.
\end{enumerate}

In the temperature chaos scenario, spin orientations are spatially uncorrelated
beyond an overlap length, $\ell_o(T_1 - T_2)$, 
which diverges when $T_1 - T_2 \to 0$ but is small
at large temperature differences, $\ell_o \sim |T_1 - T_2|^{-1/\zeta}$, 
where $\zeta$ is an exponent quantifying chaos. 
Chaos can therefore naturally explain 
full rejuvenation if $\ell_o < \xi_2$ because 
equilibrium states at the two temperatures are uncorrelated
at the important length scale of $\xi_2$.
As a consequence,
aging at $T_1$ is unable to bring the system 
closer to equilibrium at $T_2$. 
In the opposite
situation, $\ell_o > \xi_2$,
called `weak chaos', a smaller
but non-zero rejuvenation signal 
should be observed~\cite{yosh3}, since there are some rare
regions of space, occurring with probability
$(\xi_2 / \ell_o)^\zeta$, which are affected by the temperature change. 

In the two-dimensional non-disordered
XY model
below its Kosterlitz-Thouless transition temperature $T_{\rm KT}$,
equilibrium spin correlations decay with
a power of the distance $r$, $C(\mathbf{r})
\sim r^{-\eta(T)}$ where the exponent
$\eta(T)$ depends continuously on $T$. 
This is because there is a line of critical
points for $0 < T \le T_{\rm KT}$.
Hence, when temperature is changed from $T_1$ to $T_2$ 
all length scales have to readjust to the new critical 
point~\cite{mauro}, whatever 
the time spent at $T_1$. Interestingly this produces a rejuvenation 
signal which is not full but becomes gradually
larger when $T_1-T_2$ is increased~\cite{surf}. 

We now investigate which of these two scenarios corresponds most closely to
our simulations.
In Ref.~[\onlinecite{I}] we showed that spatial spin glass
correlations at short distances, $r < \xi(T,t)$, 
are well-described by algebraic decays with a temperature dependent exponent, 
$C_4(\mathbf{r}, t) \sim
r^{-\alpha(T)}$ (where $C_4$ is defined in Eq.~(\ref{C4}) below),
just as in the two-dimensional XY model.
Moreover, the rejuvenation effect we observe in the present study
has the same characteristics as the one found in the 
XY model, since it is never full but becomes gradually
larger when $T_1-T_2$ is increased. As for the Ising spin glass
model~\cite{BB1} this shows that the rejuvenation signal
observed in simulations is largely due the temperature dependence
of spatial correlations via the exponent $\alpha(T)$.  

It remains to ask, though, whether there are additional
rejuvenation effects observed in our simulations due to temperature chaos.
In Refs.~[\onlinecite{BB1,BB2}], it was argued that the answer for 
three and four dimensional Ising spin glasses is `no'. The reason is that
chaotic effects with temperature are very hard to 
detect presumably because the overlap length
is never small in Ising systems~\cite{timo,hajime}. 
However, equilibrium calculations
for vector spins suggest that overlap lengths should be smaller
in the Heisenberg case~\cite{kr}, while
dynamic length scales are larger
than in the Ising model~\cite{I}. Hence it appears potentially easier to
detect chaotic temperature effects in Heisenberg systems than
in Ising systems~\cite{II}, so this possibility needs to be investigated.

To do so we follow
Ref.~[\onlinecite{BB1}] and consider a two-site, two-replica correlation 
function,
\begin{equation}
C_4(\mathbf{r}, t) = {1 \over L^3} \sum_i
\langle \mathbf{S}_i^a(t) \cdot \mathbf{S}_{i + \mathbf{r}}^a(t) \ 
        \mathbf{S}_i^b(t) \cdot \mathbf{S}_{i + \mathbf{r}}^b(t) \rangle
\label{C4}
\end{equation}
where the two replicas $a$ and $b$ have the same interactions but
age independently 
at different temperatures, $T^a$ and $T^b$, with $\Delta T \equiv T^a - T^b$. 
If $\ell_o(\Delta T) < \xi(T^a, t)$ then the overlap between the spin
configurations in the two replicas
becomes small in a simulation on timescale $t$.

\begin{figure}
\begin{center}
\psfig{file=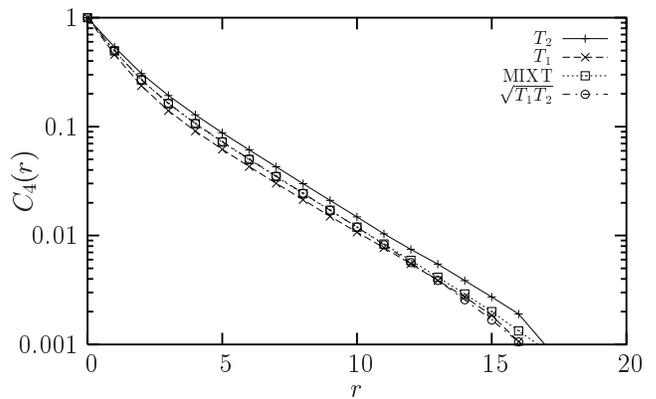,width=8.5cm}
\caption{\label{chaos} Spatial correlations
between two replicas, $a$ and $b$, at equal temperatures
$T_1=0.14$ (``$T_1$'') and $T_2=0.04$ (``$T_2$''), 
or different temperatures (``MIXT''). 
The curve noted ``$\sqrt{T_1 T_2}$'' is the 
square root of the product of the data at $T_1$ and $T_2$
and coincides with the ``MIXT'' points, as expected in the absence of
chaos.} 
\end{center}
\end{figure}

In  Fig.~\ref{chaos}
we show our results for the Heisenberg model
with the same two temperatures $T_1 =0.14$ and $T_2 = 0.04$ 
used in our previous cycles.
Data is presented for the two-replica correlators
in three different cases.
\begin{itemize}
\item
$T^a=T^b=T_1$, and $\tw^a=\tw^b=3728=t_1$, noted ``$T_1$''.
\item $T^a=T^b=T_2$, and $\tw^a=\tw^b=57797=t_2$, noted ``$T_2$''.
\item $T^a=T_1$, $T^b=T_2$, $\tw^a=3728$, and $\tw^b=57797$,
noted ``MIXT''.
\end{itemize}
The simulation
times have been chosen so that $\xi(T_1,t_1) \simeq \xi(T_2,t_2)$.
This equality can be seen in Fig.~\ref{chaos} because
the correlation functions ``$T_1$'' and ``$T_2$'' are parallel in this
log-lin representation
at large $r$. By contrast these
curves have a different slope at small $r$ because 
short distance behavior is characterized by 
the temperature dependent exponent $\alpha(T)$
discussed above.

Since the temperatures $T_1$ and $T_2$
are very different, we need to take into account the
overall increase of spin glass order as the temperature is 
lowered~\cite{yosh5_1}. We therefore also plot
$\sqrt{C_4(r,T_1,t_1) C_4(r,T_2,t_2)}$, which is
called  ``$\sqrt{T_1 T_2}$'' in Fig.~\ref{chaos}.
We find that 
``MIXT'' and ``$\sqrt{T_1 T_2}$'' are equal within our numerical
precision, the result expected in the absence of chaos.
We conclude that even weak chaotic 
effects are not detectable in our numerical simulations
despite large temperature jumps, $\Delta T / T_c \sim 0.625$, 
and large dynamic lengthscales, see the range for $r$ in Fig.~\ref{chaos}.
The dynamic rejuvenation effects we observe can therefore 
be attributed entirely to the temperature dependence of the exponent
$\alpha(T)$
describing the 
power law decay of spatial correlations at distances less than
$\xi(T, t)$. 

Of course temperature chaos could still exist at larger
length scales. Although the timescales of experiments (relative
to the microscopic time) are much longer than in simulations, the difference
in length scales is not so pronounced, as we shall emphasize in
Sec.~\ref{discussion}. Hence we feel it is unlikely that 
strong temperature chaos
occurs in experiment either.
Unfortunately, this is difficult to check since
direct probes of length scales related to chaos effects are not experimentally
feasible.
As discussed in detail in Sec.~\ref{disc:fullrejuv}, the experimental
observation of full rejuvenation does not in itself
prove that a strongly chaotic situation is 
reached in experiments.

\subsection{What is a direct quench in experiments?}
\label{direct}

To our knowledge, full rejuvenation has never been found
in numerical studies even in large temperature shifts, in contrast to 
experiments which do find full rejuvenation. One difference in procedures is
that temperature changes in simulations are instantaneous, whereas they can
only take place at a finite rate in experiments. We therefore need to study
whether correlations built up during gradual cooling are relevant to the
subsequent dynamics. If chaos were important in the simulations, the answer
would presumably be no, since growth of correlations with time has to restart
from scratch at each temperature. However, as we showed in Sec.~\ref{secchaos},
no chaos effects are seen in our simulations. Hence we need to
consider the effects of gradual changes in
temperature, to see whether these modified protocols improve agreement with
experiment.

\begin{figure}
\begin{center}
\psfig{file=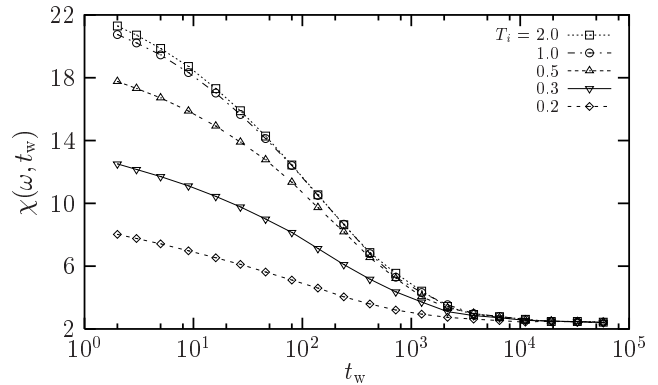,width=8.5cm}
\caption{\label{quench}
Effect of a non infinite initial
temperature in a quench. The system is first 
quenched from $T=\infty$ to various temperatures, $T_i > 
T_c \simeq 0.16$. 
Temperature is kept constant at $T_i$ during 
$10^4$ time steps after which it is quenched 
to the final temperature $T_1=0.04$ at time $\tw=0$.}
\end{center}
\end{figure}

Firstly, rather than quenching from $T=\infty$ to $T_1$, we consider
the effect of taking a finite initial 
temperature, $T_c < T_i < \infty$.
This protocol allows us to investigate the 
effect of building spatial correlations at 
high temperatures on the aging dynamics at lower temperature.
Our results for various $T_i$ are
presented in Fig.~\ref{quench} where
we have adopted the following procedure. The temperature is 
quenched from $T=\infty$ to $T_i$ where it is kept constant
for $10^4$ sweeps after which the temperature is instantaneously 
quenched to the final temperature $T_1=0.04$. 

For initial temperatures
larger than $T_i \simeq 2.0 \simeq 12.5 T_c$ the aging observed 
at $T_1=0.04$ is similar to the one observed in a quench
from $T_i=\infty$. However, as soon as $T_i < 2.0$ aging
is affected. In particular the total amplitude of 
the relaxation, 
\begin{equation}
\Delta \chi(\omega) = \chi(\omega, t=0) -
\chi(\omega, t=\infty),
\label{amplitude}
\end{equation} 
is greatly reduced when $T_i$ decreases.
We then use Eq.~(\ref{amplitude}) to 
compare the aging dynamics in various
protocols. From Fig.~\ref{rejuv} we find that
the ratio of the amplitude of relaxation in a truly direct
quench to the amplitude in a $0.14 \to 0.04$ shift 
is about 5 while full rejuvenation would imply that this ratio 
is 1. If, however, one compares the $0.14 \to 0.04$ shift 
to a direct quench from a finite temperature, 
say from $T_i = 0.3 \simeq 1.9 T_c$, to $T=0.04$, 
this ratio becomes $\simeq 2.8$, i.e.~closer 
to the full rejuvenation result.
We conclude that spending some time at high
temperatures
makes the rejuvenation signal closer to what is observed experimentally,
though we still do not see full rejuvenation.

In real experiments, not only is the initial temperature finite,
as above, but also
the cooling from $T_i$ to $T_1$ is not instantaneous. 
Therefore not only does the system build correlations
at high temperature $T_i$,  but it also has some dynamics
at all temperatures intermediate between $T_i$ and $T_1$.
This is likely to influence even more
the aging dynamics at the final temperature. 

To investigate this point 
we have performed simulations of temperature cycles 
using finite cooling rates. We repeated the 
temperature cycles of Sec.~\ref{introduction} using a 
realistic protocol. The system is first quenched from $T=\infty$ to
$T_i=2.0$ where it is kept constant during $10^4$ sweeps.
Then the temperature is decreased at a finite 
cooling rate, $R$, so that the time evolution of the 
temperature is of the form $T(t) = T_i - R t$. 
The cooling is stopped at $T_1=0.14$ for a 
time $t_1$, after which the temperature is decreased to $T_2=0.04$
at the same cooling rate $R$. 
We simulated cycles with two different cooling rates, $R_1 = 0.002$ and 
$R_2 = 0.0005 = R_1/4$ (cooling rates are expressed in units
of $J/t_0$ where $J=1$ is the variance of the distribution
of coupling constants in Eq.~(\ref{ham}) and $t_0=1$ is the 
Monte Carlo time unit). These values were chosen to give 
cooling times comparable to $1/\omega$, as in
experiments~\cite{india}.
We find that a finite cooling rate has two effects:
\begin{enumerate}
\item
All the amplitudes of relaxation are reduced. This
is a somewhat natural observation.
\item
A less trivial result is that
ratio of amplitudes between a direct quench to $T_2$
and a shift through $T_1$ are again reduced from 
their $R = \infty$ values. When $1/\omega=1078$, we find 
that these ratio are 5, 3.8, and 3 for $R=\infty$, 
$R_1$ and $R_2$, respectively.
\end{enumerate}
These results again indicate 
that the finite cooling rates used in experiments, 
make the observed rejuvenation 
effects closer to that in
experiments, though we are still far from
full rejuvenation. In fact, we shall argue in Sec.~\ref{disc:fullrejuv}
that full rejuvenation is
expected in experiments because of the much larger range of timescales that
are probed there than in the simulations.

\begin{figure}
\begin{center}
\psfig{file=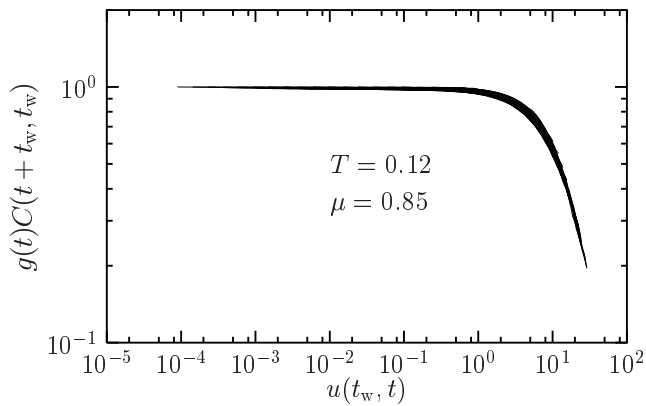,width=8.5cm}
\caption{\label{sub} Subaging behavior
found 
at $T=0.12$ when the initial temperature is 
$T_i=0.3$ and the cooling rate infinite. 
The horizontal axis is 
$u(\tw,t) = [(t+\tw)^{1-\mu}-\tw^{1-\mu}]/(1-\mu)$ which 
reduces to $t/\tw$ when $\mu=1$, and the vertical axis involves a factor
$g(t)$ to account for quasi-equilibrium short-time behavior,
see Ref.~[\onlinecite{I}] for details. The parameter $\mu$ is adjusted to get
the best data collapse for different values of $\tw$.
The case $\mu < 1$, as found here, corresponds to subaging. The
case, $\mu=1$, which we find when
the initial temperature is $T=\infty$, corresponds to 
simple aging.}
\end{center}
\end{figure}
  
We have shown that a 
non ideal quench very strongly affects aging behavior. 
In fact we find that all scaling behaviors reported 
in Ref.~[\onlinecite{I}] for simple aging experiments
are affected. This is a likely explanation of the `sub-aging' behavior
systematically found in 
experiments~\cite{review1,india} while simulations
indicate instead `super-aging' behavior~\cite{I,BB1}. As a single 
example we show in Fig.~\ref{sub} the effect of a non-infinite 
initial temperature, $T_i = 0.3$, on the aging 
at temperature $T=0.12$. The cooling rate is infinite.
When $T_i = \infty$ a simple scaling of correlation
functions is observed at this 
temperature~\cite{I}, $C(\tw, t+\tw) \sim g^{-1}(t) {\cal C}(t/\tw)$, 
where $g(t)$ accounts for the short-time quasi-equilibrium behavior.
Instead, when $T_i=0.3$, we find that a more complicated, sub-aging scaling
form is needed, as shown in Fig.~\ref{sub}.  
We have 
systematically found sub-aging behavior
in non-ideal quenches.
Given that experimental cooling times are always extremely large 
when compared to microscopic timescales it is 
not surprising that this effect persists 
experimentally~\cite{india}, unless specific but empirically
determined protocols are used~\cite{cool}. 
Cooling rate effects also imply that aging 
dynamics probed (a) numerically, and (b) experimentally in `direct 
quenches', are fundamentally different, which prevents a
quantitative comparison of their scaling behavior.

\section{Memory}
\label{memory}

Memory effects are simpler to understand
than rejuvenation effects. They directly result
from the strong influence of temperature on the timescales needed to
equilibrate fluctuations on a given length scale.
Dynamics at different temperatures
but in similar time windows probe different length 
scales~\cite{review3,BB1,fh,jp,encorejp,yosh5}.
In this section we quantify these effects. 

\subsection{Is there perfect memory for large shifts?}

\begin{figure}
\begin{center}
\psfig{file=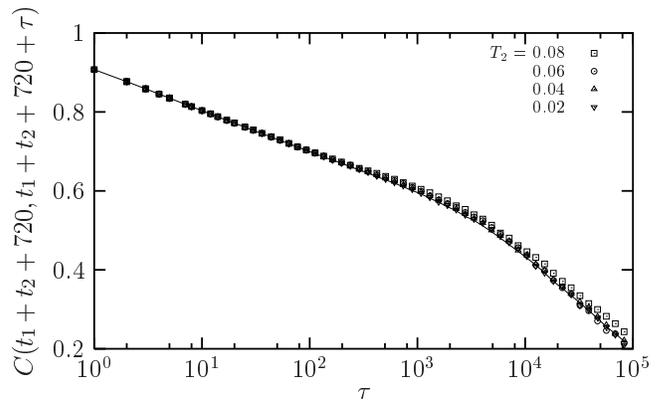,width=8.5cm}
\caption{\label{memoryfig} Time decay of the autocorrelation function
$C(t_1 + t_2+720, t_1 + t_2 + 720 + \tau)$ in the third step of temperature 
cycles with perfect memory.
The protocol is as in Fig.~\ref{cycle}, i.e.~$T_1=0.14, T_2 = 0.04, t_1 = t_2
= 11159$. The waiting time $\tw = t_1 + t_2 + 720$
corresponds roughly to the end of the 
transient seen in Fig.~\ref{cycle}. 
The full line is the autocorrelation measured in a direct quench to 
$T_1=0.14$ after waiting $t_1=11159$,
i.e.~ $C_{T_1}(t_1, t_1 + \tau)$. If memory is perfect, the data 
from the cycle agrees with that from the direct 
quench. Small deviations from perfect memory
are visible at $T_2 = 0.08$, but absent for lower $T_2$ 
for which `memory's perfect'.}
\end{center}
\end{figure}

In the inset of Fig.~\ref{cycle} we have shown that 
the aging in the third step of the temperature cycle
appears to be, after a short transient, 
the perfect continuation of the first step. However, this data is
not completely convincing because 
$\chi(\omega,\tw)$ decreases very slowly on the timescale of the 
simulation.  It is
better to study the full frequency dependence, or, equivalently, the 
time dependence of the 
spin autocorrelation function~\cite{BB1}.

In Fig.~\ref{memoryfig} we show that the memory effect observed 
in Sec.~\ref{introduction} is indeed convincing, or, 
in an allusion to Billy Wilder, that `memory's perfect'~\cite{www}. 
The decay of the autocorrelation function in the third stage of
the cycle for large $T_1 - T_2$ is indeed the same,
within our numerical accuracy, 
as the one obtained if the second stage of the experiment is dropped out.
As in experiments, we conclude that despite the strong rejuvenation
effect observed in the second stage at temperature $T_2$,
the correlations built in the first stage at temperature $T_1$
have been perfectly preserved and can be retrieved when 
temperature is shifted back to its initial value.

\subsection{Is there memory for small shifts?}
\label{memory_heis}

In the previous subsection we showed that memory is essentially perfect for
large temperature shifts. We now discuss what happens for smaller values of
the temperature shift where we will see that cumulative aging becomes
a good approximation. To compare with the cumulative aging scenario
we extract an effective age $\teff$ by
fitting the data for $\chi(\omega, t_1+t_2 + t)$ to Eq.~(\ref{chimemstepup}).
Ideally one can find a $\teff$ value so that this data is equal to
$\chi_{_{T_1}}(\omega, t_1 + \teff + t)$ for all $t$. 
For technical reasons it is difficult to measure the effective age 
when times get large because the change in $\chi$ becomes too small.
Hence we will extract $\teff$ using a protocol with $t_1=0$, i.e.~a
sudden quench to $T=T_2$ followed by an upward shift to $T=T_1$ at time $t_2$.
This modified protocol does not correspond to 
the cycle discussed in Sec.~\ref{introduction} but is used here
as a  convenient mean to measure effective aging times. Also
we will actually do the fit to the time dependent correlation
function in Eq.~(\ref{sus}). This is, of course,
equivalent to fitting data for $\chi$.

\begin{figure}
\begin{center}
\psfig{file=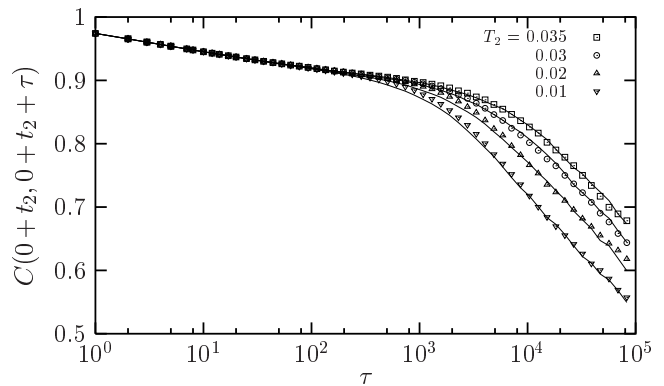,width=8.5cm}
\caption{\label{teff1} Influence of the aging at low temperature
$T_2 < 0.04$ on the subsequent aging at $T_1=0.04$. The system is
quenched directly from $T=\infty$ to $T_2$ (so $t_1 = 0$),
where it is kept during
$t_2= 11159$ time steps, after which the temperature is shifted up to 
$T_1=0.04$. Autocorrelation functions in these shifts are 
shown with symbols. They are compared
to the ones, shown with full lines,
measured in a direct quench to $T_1=0.04$ and different waiting times
$\tw=2700$, 4500, 7000 and 9500 (from left to right), chosen to match 
the temperature shift data. The matching is seen to work well.}
\end{center}
\end{figure}

Figure~\ref{teff1} shows some data for this protocol 
as a function of $\tau =
\omega^{-1}$ with $T_1 = 0.04$,
$t_1 = t = 0$, and $t_2 = 11159$.
Just after the shift up to $T_1$ we record the time 
decay of the spin autocorrelation functions. By definition, 
when $T_2=T_1$ the autocorrelation function follows
the curve obtained in a direct quench to $T_1=0.04$
after $\tw = 11159$. However,
when $T_2 < T_1$ the time decay of the autocorrelation
function after the shift becomes faster, but 
an effective age, $\teff < \tw$, can still be defined, 
since the solid lines
in Fig.~\ref{teff1}, which are from direct quenches, go through the points,
which are from the temperature shift protocol. In other words, matching the
correlation functions
as a function of $\tau=\omega^{-1}$ to the expression analogous to
Eq.~(\ref{chimemstepup}) with $t_1=t=0$ works well.
We see from Fig.~\ref{teff1} that
$\teff$ progressively decreases when $T_2$ decreases, as 
expected in the cumulative aging scenario of Sec.~\ref{cumag}. 
Physically this means that aging at low temperature is less 
effective at building spatial correlations, or, 
in other words, that the growth of the dynamic correlation length
is slower at lower temperature.

\begin{figure}
\begin{center}
\psfig{file=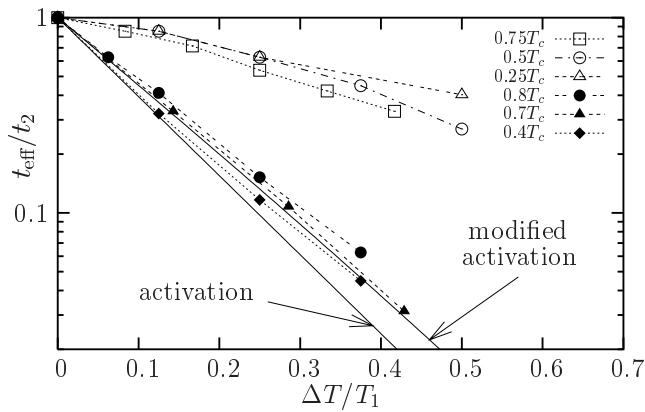,width=8.5cm}
\caption{\label{teff} 
The effective age $\teff$ in a shift experiment for different values of $T_1$
at fixed $t_2$. The system is quenched 
at $T_2 = T_1 - \Delta T$ where it ages during a time $t_2=11159$ after which
the temperature is shifted up to various temperatures $T_1$.
Open symbols are for the Heisenberg spin glass, filled symbols
for the Ising spin glass.
The solid line marked ``activation'' is the prediction from cumulative aging
assuming barrier activation, Eq.~(\ref{logteffcum}).
The line marked ``modified activation'' is also for the
cumulative aging scenario but assuming different form for $\xi(T, t)$, namely
$\xi(T,t) \sim t^{1/z(T)}$
with $z(T) = (T_0/T)^{0.85}$, which fits growth laws 
for the Ising spin glass.}
\end{center}
\end{figure}

\begin{figure}
\begin{center}
\psfig{file=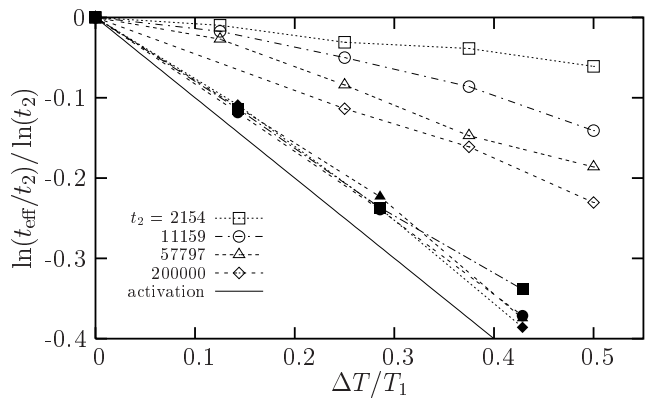,width=8.5cm}
\caption{\label{tefft}
Effective ages in shift experiments for different values of $t_2$ at fixed
$T_1$. The system is quenched 
at $T_1 - \Delta T$ where it ages during a time $t_2$. 
The temperature is then shifted up to $T_1$.
Open symbols are for the Heisenberg spin glass
with $T_1=0.08 \simeq 0.5 T_c$, 
filled symbols for the Ising spin glass with $T_1=0.7 T_c$
but the same times $t_2$ as the Heisenberg model.
The straight line is the prediction of cumulative aging with
thermal activation, Eq.~(\ref{logteffcum}).}
\end{center}
\end{figure}

We have repeated this matching
procedure for many different shifts, changing
$t_2$ from 2154 to $2 \times 10^5$ time steps, 
the final temperature from $T_1=0.4$ to $T_1=0.12$, 
and the amplitudes of the shift, $\Delta T = T_1-T_2$, from
$\Delta T=0.01$ to $\Delta T=0.05$. The results are 
presented in Figs.~\ref{teff} and \ref{tefft} which
also contain similar data for the 
Ising spin glass that are discussed below in 
Sec.~\ref{anisotropy}.

\begin{table}[b]
\caption{\label{table} 
Evaluation of the cumulative aging hypothesis 
in small temperature shifts.}
\begin{ruledtabular}
\begin{tabular}{c c c c c c}
$T_2$ &  $T_1$ & $t_2$ & $\teff$ & $\teffcum$ & $\teff / \teffcum$ \\
\hline
0.02 & 0.04 & 11159 & 4500 & 4200 & 1.07 \\
0.04 & 0.08 & 11159 & 3000 & 4200 & 0.71 \\
0.10 & 0.12 & 11159 & 8000 & 5500 & 1.45 \\
0.08 & 0.12 & 11159 & 4700 & 3800 & 1.24 \\
0.04 & 0.08 & 2154  & 1350 & 1300 & 1.04 \\
0.04 & 0.08 & 57797 & 7500 & 11000 & 0.68 \\
0.10 & 0.12 & 57797 & 29000 & 15000 & 1.93 \\
0.08 & 0.12 & 57797 & 13000 & 11500 & 1.13 \\
0.14 & 0.12 & 11159 & 22500 & 22000 & 1.02 
\end{tabular}
\end{ruledtabular}
\end{table}

We now compare our results for $\teff$ to the prediction of the cumulative
aging scenario
\begin{equation}
\xi(T_1, \teffcum ) = \xi(T_2, t_2) \, ,
\label{xi2}
\end{equation}
which is just Eq.~(\ref{xi1}) with $t_1 = t' = 0$. 
For each of the parameters $T_1, T_2$ and $t_2$ corresponding to the
points for the Heisenberg model in Fig.~\ref{teff}, we determined $\teffcum$
using Eq.~(\ref{xi2}) and data for $\xi(T,t)$ in Ref.~[\onlinecite{I}].
The ratios of $\teffcum$ to
$\teff$ determined from fits like those in Fig.~\ref{teff1}
are presented in Table~\ref{table}.
We see that the ratio roughly
takes values in the range $[\frac{1}{2}, 2]$, 
with no obvious systematic behavior.
That the ratio is not perfectly 1 is hardly surprising given that 
one has to make use twice of imperfect numerical data to estimate
$\teffcum$ from Eq.~(\ref{xi2}). 
We conclude, contrary to the preliminary investigations 
of Ref.~[\onlinecite{ricci2}] that the cumulative aging 
hypothesis, Eq.~(\ref{xi2}), is a good 
interpretation of the behavior of the Heisenberg spin 
glass in temperature shifts of small amplitude. 
Quite importantly, these results validate the
assumption of cumulative aging used in
Refs.~[\onlinecite{dupuis,bert,encorejp}], 
where experimental data plotted as in
our Figs.~\ref{teff} and \ref{tefft} is one of
the ingredients
used to reconstruct growth laws for $\xi(T,t)$.

Alternatively, rather than estimate $\xi(T, t)$ from the numerical data of
Ref.~[\onlinecite{I}], one can assume an analytical form for the growth law,
for example the barrier activation model used to obtain Eqs.~(\ref{T1lnt1}),
(\ref{T2lnt2}) and (\ref{T2lnt'}). This predicts
\begin{equation}
\ln \left({\teffcum \over t_2}\right)
= - \frac{\Delta T}{T_1} \ln t_2 \, ,
\label{logteffcum}
\end{equation}
which is equivalent to Eq.~(\ref{T1lnt1}) with the roles of $T_1$ and $T_2$
interchanged. It corresponds to the straight solid line marked ``activation''
in Figs.~\ref{teff} and \ref{tefft}.
We find that for all temperatures the 
measured $\teff$ lie significantly above this result.
Moreover, the dependence upon the final
temperature $T_1$ at fixed $t_2$ is weak.
From the growth laws for $\xi(T,t)$ reported in Ref.~[\onlinecite{I}]
this result is reasonable, since the dynamic correlation
length is very weakly dependent on temperature for times 
smaller than $t \sim 10^3$ and only after that does it enter
a regime dominated by thermal activation~\cite{I}. 
We therefore expect that shifts performed at larger 
times $t_2$ will get closer to the cumulative aging scenario with a
barrier activation form for $\xi(T, t)$.

We confirm this expectation 
in Fig.~\ref{tefft} where we work with a constant final temperature, 
$T_1 = 0.08 \simeq 0.5 T_c$, but change the duration
of the initial period over two orders of magnitude.
To compare data with different shifting times, we 
plot $\ln(\teff/t_2) / \ln t_2$ vs. $\Delta T / T_1$
for which cumulative aging with
thermal activation, Eq.~(\ref{logteffcum}), predicts a linear relation
of slope $-1$, independent of $t_2$.
Upon increasing $t_2$ we find that the data get closer to this prediction,
but still lie systematically above it. 
Similar results are found for $T_1 = 0.12 \simeq 0.75 T_c$.
As a rough estimate, if the evolution of these curves 
persists at much larger times
it would require about 4 more orders of magnitude, 
i.e. $t_2 / t_0 \sim 10^9$, to get curves that 
eventually lie on or below 
the thermally activated estimate. This is not inconsistent 
with the experimental finding that these curves 
lie below the Arrhenius line in Heisenberg spin glass 
samples~\cite{dupuis,bert,india}
since experiments are performed on timescales that 
are typically $10^{12}$--$10^{16}$ times larger than 
microscopic flipping times,
see e.g.~Fig.~\ref{th}.

In this section we performed 
upward temperature shifts of relatively small amplitude.  
We expect that cumulative aging would also work for small temperature changes
in a downward shift, but for
technical reasons it is difficult to test extensively this
hypothesis directly, because effective ages
very quickly become too large to be numerically measurable
in downward shifts. We nevertheless add in Table~\ref{table} the result
for a downward $0.14 \to 0.12$ shift from the data
displayed in Fig.~\ref{rej}a. For this particular example, 
cumulative aging works indeed very correctly.

\subsection{Does spin anisotropy play a role?}
\label{anisotropy}

In this subsection we compare our results for the temperature shifts
with similar results obtained in the three-dimensional 
Ising spin glass, using exactly the same 
thermal protocols and the same procedure to extract effective ages.
To this end we needed to extend
significantly the numerical results 
obtained by one of us and Bouchaud in Ref.~[\onlinecite{BB1}]
to which we refer for technical details concerning
the Ising simulations. 
These new results for the three dimensional Ising Edwards-Anderson 
spin glass are reported along the ones for the Heisenberg model 
in Figs.~\ref{teff} and \ref{tefft}. 

A quantitative comparison between the two sets of data 
can be performed if one assumes that in both cases, 
Monte Carlo time units represent the microscopic 
flipping time which we fix to $t_0=1$ in both 
simulations---a physically reasonable assumption.
From the comparison between the two models one
can draw two main conclusions.

\begin{itemize}

\item 
Curves for the Ising spin glass lie below the ones
of the Heisenberg model for all the shifts we have performed, so 
$\teff/t_2$ is smaller for the Ising case. In other words, the effect of a
temperature shift on aging dynamics is stronger for the Ising simulations.
However, the opposite behavior is observed in 
experiments which compare strongly anisotropic (Ising) samples with 
isotropic (Heisenberg) samples~\cite{bert}.

\item 
Changing the duration of the shift over two orders of magnitude 
has a strong effect on the Heisenberg spin glass, as described in
Sec.~\ref{memory_heis}
above, but very little or no effect in the Ising case, see Fig.~\ref{tefft}.
To our knowledge there are no systematic
experimental investigations concerning this point but 
if this trend persists to much larger timescales, it could lead to a situation
where Heisenberg data lies below the ``activation'' curve while Ising data
lies above it, as is found experimentally~\cite{bert}.

\end{itemize}

The Ising data agrees better with the barrier activation form for $\xi(T,
t)$ in the cumulative aging picture than does the Heisenberg data. Although we
use the terminology ``barrier activation'' the same result,
Eq.~(\ref{logteffcum}),
applies if the
barrier height only depends logarithmically on length scale,
in which case~\cite{rieger,yosh2,ricci2}
$\xi(T, t) \sim t^{1/z}$ with $z = T_0/T$, i.e.~$\xi$ grows 
with a power of $t$.
Recently, algebraic laws with $z = (T_0/T)^\gamma$
with $\gamma \simeq 0.85$ were
reported~\cite{ricci2}. 
As shown in Fig.~\ref{teff} the introduction of the 
adjustable parameter $\gamma$ allows one to
understand deviations from simple thermal activation 
in the Ising spin glass within the 
cumulative aging hypothesis. 
Possibly critical fluctuations near $T_c$
renormalize the microscopic flipping time~\cite{dupuis,yosh4,BB1}
in such a way that the growth is approximately of this form
in numerical time windows.
It would be interesting to check if experimental data
taken in anisotropic (Ising)
samples can be fitted with the same 
``modified activation'' growth law.  

\section{Discussion}
\label{discussion}

\subsection{Memory effects}
\label{disc:memory}

Memory effects stem from thermally activated dynamics which implies 
that temperature so strongly influences the growth of the 
dynamic correlation length, $\xi(T,t)$, 
that aging is much less effective at building
spatial correlations at lower temperatures. 
Therefore lengthscales that are quasi-equilibrated 
at some temperature $T_1$ can effectively be frozen if the temperature
is decreased to $T_2< T_1$. Spatial patterns imprinted 
in the systems at $T_1$ are then naturally retrieved when temperature is 
shifted back to its original value. 
From this perspective it is natural that memory effects 
are found in so many different disordered 
materials where activated dynamics is 
ubiquitous~\cite{I,review1,review1_1,review3,dupuis,yosh4,yosh3,bert,cycle,physica,dip,petra,kovacs,india,matthieu,struik,frustre,leheny1,leheny2,beta,KLT,bellon,ferro,ferro2,electron}.

\begin{figure}
\begin{center}
\psfig{file=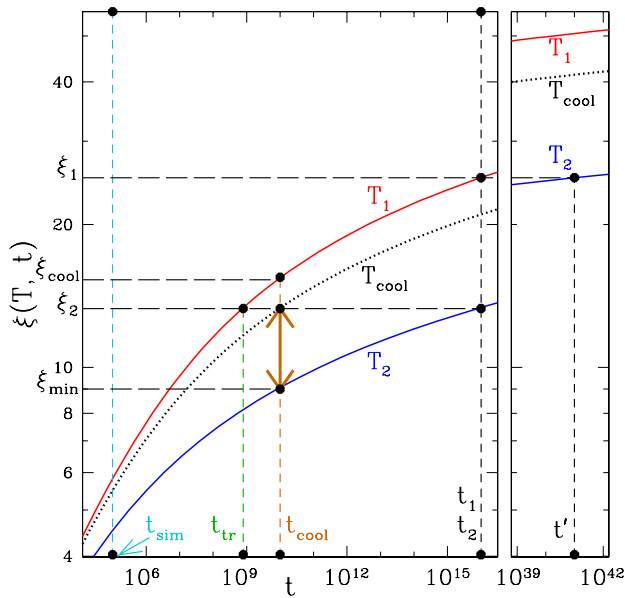,width=8.5cm}
\caption{
\label{th}
Solid curves show $\xi(T, t)$ (with $t$ in units of the microscopic time
$t_0=10^{-12} \, s$) inferred in Ref.~[\onlinecite{encorejp}] for 
a Heisenberg spin glass 
at temperatures $T_1/T_c=0.825$,
and
$T_2/T_c=0.7$. Note the break in the
horizontal scale between the two boxes.
We assume waiting times at $T_1$ and $T_2$ of $t_1 = t_2= 10^4\, {\rm s}
=10^{16} t_0$.
The length scales equilibrated at $T_1$ and $T_2$ are 
$\xi_1 \equiv \xi(T_1, t_1)$ and
$\xi_2\equiv \xi(T_2, t_2)$ (indicated by
horizontal long-dashed lines).
To reach length scale $\xi_1$ at the lower temperature $T_2$ one
would have to wait a time $t'\simeq 10^{41}\, t_0 \simeq 3 \times 10^{21}
\, {\rm years}$, 
given by Eq.~(\ref{xi'}).
This is truly astronomical, so
fluctuations between $\xi_2$ and $\xi_1$ are effectively frozen during the time
$t_2$ spent at $T_2$, which is the origin of the memory effect.
When the temperature is raised
back to $T_1$, see Fig.~\ref{cycle}, fluctuations on length scale $\xi_2$ have
to reequilibrate at $T_1$, which takes a ``transient''
time $t_{\rm tr} \simeq 10^9 =
10^{-3} \, {\rm s}$ (which is very short;
another requirement for perfect memory).
We also show $\tcool$, the time spent `close to' a given
temperature during cooling,
and indicate, by a dotted line,
the growth at $T=\Tcool = 0.795 T_c$, the temperature at which
$\xi(\Tcool, \tcool)$ is equal to $\xi_2$. We also define
$\xicool=\xi(T_1, \tcool)$. In cooling to $T=T_2$ the system will have
correctly equilibrated for temperature $T_2$ up to scale
$\ximin=\xi(T_2,\tcool)$. During
subsequent
aging for time $t_2$ at $T=T_2$, fluctuations on scales between $\ximin$ and
$\xicool$ (shown by the thick line with arrows)
have to reequilibrate, which gives rise to rejuvenation.
The time scale probed in our
simulations is of order $t_{\rm sim} \simeq 10^5$, which is
much smaller than in experiments.
However, the difference between the length scales probed in experiment and
simulations is much smaller than the difference in timescales.}
\end{center}
\end{figure}

In Fig.~\ref{th} we show an example of
growth laws as extracted from experimental 
data in Heisenberg samples in Ref.~[\onlinecite{encorejp}]. The precise
details of the fits which gave these curves will not be relevant here because
our main conclusions will be qualitative.
We use this graph to discuss a temperature 
cycle occurring between temperatures $T_1 = 0.825 T_c$ and $T_2 = 0.7 T_c$. 
With\cite{encorejp} $T_c = 16.7 \, {\rm K}$, this gives $T_1 = 13.77 
\, {\rm K}$ 
and $T_2 = 11.69 \, {\rm K}$, so that $\Delta T \simeq 2 \, {\rm K}$, 
as in the original experiment of Ref.~[\onlinecite{cycle}].

For concreteness we assume a microscopic time of $t_0=10^{-12} \, {\rm s}$
and take $t_1 = t_2 = 10^4\, {\rm s}$ so, in units of the microscopic time, we
have $t_1 = t_2 = 10^{16}$. The correlation lengths
$\xi_1$ and $\xi_2$ reached at $T_1$ and $T_2$ in a time 
$t_1=t_2$ are 25.3 and 13.3 respectively,
indicated by dashed horizontal lines. To reach a length scale of $\xi_1$
at the lower temperature $T_2$
would require an astronomically large time of order $t' \simeq 10^{41}\,
{\rm t_0} = 10^{29}\,{\rm s}$,
given by Eq.~(\ref{xi'}), which is indicated in the right hand
box of the figure.
Also relevant is an estimate of 
the transient time, $t_{\rm tr}$, required
to reequilibrate fluctuations at scale $\xi_2$ 
when the temperature is increased back to
$T_1$. This is given implicitly by
$\xi(T_1, t_{\rm tr}) = \xi_2 \equiv \xi(T_2, t_2)$.
From Fig.~\ref{th} we see that $t_{\rm tr}$ is of order $1\,{\rm msec}$,  
hardly detectable if the frequency is 
of the order of a few Hz.
In general, we expect
perfect memory if Eq.~(\ref{t'ggt2}) and the condition
\begin{equation}
t_{\rm tr} \ll t_2  \qquad \mbox{(needed\ for \ perfect\ memory)},
\label{ttr} 
\end{equation}
are both satisfied.
Within the barrier activation
model and for $t_1 \sim t_2$,
Eqs.~(\ref{ttr}) and (\ref{t'ggt2}) are actually
equivalent since this model
predicts that
\begin{equation}
{\ln t_2 \over \ln t_{\rm tr}} = {\ln t' \over \ln t_1} = {T_1 \over T_2}.
\end{equation}
Equation (\ref{memperf}),
which corresponds to Eq.~(\ref{t'ggt2}) for the barrier
activation model, is satisfied for the experimental situation in
Fig.~\ref{th}, so memory
should be perfect, as was indeed found~\cite{cycle}. 
However, Eq.~(\ref{memperf}) is
harder to satisfy 
in numerical simulations since the timescales
are shorter.
Hence larger temperature steps have to be chosen
in simulations, but, nonetheless, Eq.~(\ref{memperf})
can then be satisfied.
This conclusion holds 
for both Heisenberg and Ising spin glass 
models~\cite{heiko,yosh2,BB1,ricci,ricci2,jimenez}.

While the details of
the numbers in this discussion
can be debated, it is surely true in general that changing the
temperature in experiments gives a large
change in timescales (though not a very large change in length scales),
and that perfect memory can be
obtained in this situation with moderate to large temperature shifts.

\subsection{Physical origin of rejuvenation effects}
\label{disc:rejuv}
 
We have seen in Sec.~\ref{disc:memory}
that to get memory it is sufficient to
have a wide separation of time scales.
We now turn to rejuvenation effects which are more difficult to understand
since, by itself, Fig.~\ref{th} does not explain why 
aging is strongly restarted when the temperature is shifted from 
$T_1$ to $T_2$.
The important question is that,
since excitations on lengthscales active at $T_2$ were fully equilibrated at 
$T_1$ just before the shift, why do they 
need to reequilibrate at all upon a small temperature change?

In this paper we have been able to answer this question 
for the Heisenberg spin glass model simulated on timescales 
up to $10^5$. We find that spatial correlations
at equilibrium (i.e.~at distances less that $\xi(T, t)$)
are close to algebraic with a temperature
dependent exponent $\alpha(T)$ so that excitations on all 
lengthscales up to $\xi$ have to readapt at each temperature. 
This interpretation is very close to the physical 
interpretation of the behavior of the two-dimensional
XY model when temperature is shifted along its critical line~\cite{surf}. 

As discussed in Section~\ref{secchaos} we found no trace of chaotic behavior
of spatial correlations with temperature, though it is possible that chaotic
effects could set in at larger lengthscales.  However, in the hypothetical
temperature cycle of Fig.~\ref{th}, which corresponds to experimental
parameters, we see that the difference in lengthscales probed by simulations
and experiments is not very drastic (a factor of about 3 or 4) so we are
tempted to conclude that chaos also does not occur in experiments.
Even if it did, chaotic effects would appear in addition
to the signal characterized in this paper, so
experimental results would present a mixed character,
difficult to analyze quantitatively within one scenario or the other.

The situation found here is qualitatively very similar to what
is observed numerically in Ising spin glasses 
in various dimensions~\cite{heiko,yosh2,BB1,BB2,ricci}. 
Differences between models arise only at a quantitative level. For instance, 
in the three dimensional Ising spin glass, the exponent $\alpha$ only depends
weakly on temperature 
taking the value of about $0.5$ throughout the spin glass phase~\cite{BB1}.
Correlations change more significantly 
in four dimensions\cite{BB1} with $\alpha$ changing from 
0.9 and 1.6 between $0.5 T_c$ and $T_c$. As a result, rejuvenation effects
are relatively pronounced in four dimensions, while almost 
absent in three~\cite{yosh2,ricci}.
By comparison, $\alpha$ in the three dimensional Heisenberg spin glass
has an intermediate behavior~\cite{I} 
since it changes 
from 0.8 to 1.1 between $0.5T_c$ and $T_c$. However this 
relatively small variation is compensated by the fact that
dynamic lengthscales for the Heisenberg model 
are also larger than for the Ising case. 

\subsection{Experimental observation of full rejuvenation}
\label{disc:fullrejuv}

We have seen in Sec.~\ref{disc:rejuv} that the observation of partial
rejuvenation in the simulations for large temperature shifts can be well
understood. Here, we give a plausible explanation of the full
rejuvenation observed experimentally,
an important ingredient being a very large separation of 
time scales, much larger than is feasible in simulations.

As mentioned in Section~\ref{direct}, `direct' quenches in experiment are not
instantaneous, and incorporating this into the simulations gave results
which were somewhat closer to full rejuvenation. Here we develop
a phenomenological description of aging which includes a
finite cooling rate.
An important concept in it is the
time spent `close to' a given temperature~\cite{cooling}, 
where a precise definition
of `closeness' will not be needed to illustrate the main points
qualitatively, but presumably requires that
spatial correlations (on the cooling timescale)
do not change `much' while the temperature is `close' to a given value.
Let us denote
this timescale by $\tcool$. 
As an example, if we assume that a quench takes place in $1\, {\rm s}$, and
divide this time into 100 intervals, we have $\tcool = 0.01\,{\rm s}
= 10^{10} t_0$, as indicated in Fig.~\ref{th}.
This is certainly rough, but
we can vary $\tcool$ quite substantially without invalidating
our main conclusions. 

It will be convenient, for the
subsequent discussion, to
define various length and time scales which arise in quenching down to $T=T_2$
at a given rate.
Correlations on length scale $r$ will freeze, i.e. 
drop out of equilibrium, at temperature
$T_{\rm f}(r)$ where 
\begin{equation}
\xi(T_{\rm f}(r), \tcool) = r \,.
\label{Tfreeze}
\end{equation} 
To determine the extent of rejuvenation
in a $T$-shift to $T_2$, we need to consider the correlations on length scales
up to $\xi_2$. The temperature where correlations on this scale fall
out of equilibrium is
\begin{equation}
\Tcool = T_{\rm f}(\xi_2)\, ,
\label{Tcool}
\end{equation}
which is $\Tcool \simeq 0.795 T_c$ in the example shown in Fig.~\ref{th}.
During the cooling process, when the temperature has reached
$T=T_1$ the correlation length will
be 
\begin{equation}
\xicool \equiv \xi(T_1, \tcool) \, 
\label{xicool}
\end{equation}
see Fig.~\ref{th}.
When the system has been cooled down to $T=T_2$, the system will be
equilibrated for this temperature up to scale 
\begin{equation}
\ximin = \xi(T_2, \tcool) ,
\end{equation}
which is also shown in Fig.~\ref{th}.


We are now in a position to compare aging after a `direct quench' to $T_2$
with that following a temperature shift from $T_1$ to $T_2$. 
Suppose first that $T_1 > \Tcool$ (i.e.~$\xicool > \xi_2$),
which is the situation in Fig.~\ref{th}. 
Then the time spent waiting at $T_1$ will have no effect on
the behavior at $T_2$ because this will only change correlations at scales
larger than $\xi_2$ that are anyway frozen at $T_2$.
The important correlations are those of scales between $\ximin$ and
$\xi_2$ and these only freeze out during
the subsequent cooling from $\Tcool$ to $T_2$, see Fig.~\ref{th}.
Provided correlations on length scales between $\ximin$ and $\xi_2$ 
depend on temperature (through the temperature dependence of 
the exponent $\alpha(T)$), these will have to reequilibrate while waiting at
$T_2$, and so there will be a rejuvenation signal.
Furthermore, the signal will be same for the direct quench and the
temperature shift from $T_1$, i.e.~we have full rejuvenation.

On the other hand, if $T_1 < \Tcool$ (so $\xicool < \xi_2$)
waiting at $T_1$ does enhance
correlations at scales $\le \xi_2$ (to be precise, at scales
between $\xicool$ and
$\xi_2$) relative to a direct quench. Hence rejuvenation will only be
partial in this case.

Hence we can succintly express the condition for full
rejuvenation as
\begin{equation}
\xicool > \xi_2 
\quad \mbox{(full\ rejuvenation)} ,
\label{frcond}
\end{equation}
or in other words, 
\begin{quotation}
\noindent the correlation length developed at $T=T_1$ 
during cooling is greater than the correlation length subsequently
developed during waiting at the lower temperature $T=T_2$.
\end{quotation}

To get a rough idea of the numbers implicit in Eq.~(\ref{frcond}), we take
the simple barrier activation model for $\xi(T, t)$ discussed in
Sec.~\ref{cumag} in which $\xi(T, t)$ is assumed to be
a function of $T \ln t$. 
The condition for full rejuvenation, 
Eq.~(\ref{frcond}), then becomes 
\begin{equation}
{T_1 - T_2 \over T_2} >  {\ln(t_2/\tcool) \over \ln \tcool  }
\qquad \mbox{(full\ rejuvenation)} .
\label{T1T2}
\end{equation}
The deviations from simple barrier activation used in
the plots in Fig.~\ref{th} actually
allow full rejuvenation when $T_1$ is somewhat closer to $T_2$ than given by
Eq.~(\ref{T1T2}).
Since we need $t_2 \gg \tcool$ (otherwise the signal is
dominated by initial transients), Eq.~(\ref{T1T2}) implies that $t_2$ 
must be huge. 

Given our numerical results~\cite{I} for $\xi(T,t)$,  
we estimate that $\tcool > 10^3$ is 
at least necessary to satisfy Eq.~(\ref{frcond}), implying that the 
total cooling time is about $10^5$, so that aging times 
should be $t_1 \sim t_2 \gg 10^5$ which is 
not presently feasible in simulations. 
This justifies our inability to reach the full rejuvenation 
limit in this numerical work, as opposed to experimental investigations.
By contrast,
perfect memory effects can be obtained in simulations because they
do not rely on the existence of finite cooling rates; compare
Eqs.~(\ref{memperf})
and (\ref{T1T2}).

\begin{figure}
\begin{center}
\psfig{file=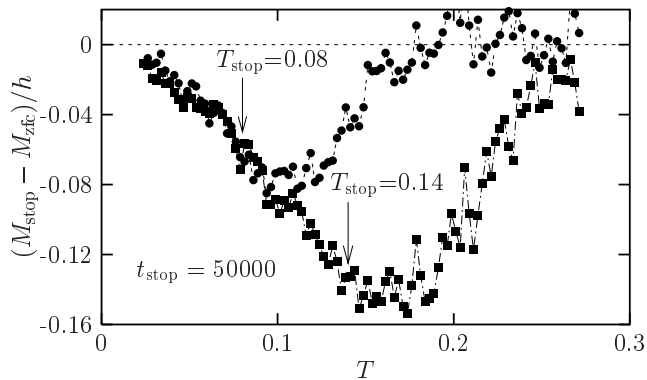,width=8.5cm}
\caption{
\label{stop}
Effect of a stop during cooling on the zero-field cooled magnetization.
The duration of the stop is $t_{\rm stop}=5\times 10^4$, 
and two temperatures $T_{\rm stop}=0.14$ (squares) and  
$T_{\rm stop}=0.08$ (circles) are used, 
as shown by vertical arrows. We plot the difference 
between $M_{\rm stop}$ and $M_{\rm zfc}$, the 
zero-field cooled magnetizations with and without the stop, respectively.
The cooling and heating rates are $R=5 \times 10^{-6}$,
the magnetic field is $h=0.01$.
A dip centered around $T_{\rm stop}$ is observed, but this is broader
than in experiments.}
\end{center}
\end{figure}

We also suspect that Eq.~(\ref{T1T2})
is necessary to observe the spectacular `dip' in ac
susceptibility measurements\cite{dip}. The reason a
sharp dip has not been seen in
simulations\cite{ricci} is, in our view,
because the timescales did not satisfy Eq.~(\ref{T1T2}).
In an effort to get closer to Eq.~(\ref{T1T2}),
we have reproduced the experimental protocol of 
Ref.~[\onlinecite{matthieu}] where a dip experiment is realized 
using the magnetization as a physical observable. 
The system is gradually 
cooled at rate $R=5 \times 10^{-6}$ from 
high temperature, $T_i = 0.35$, to very low temperature $T=0.02$.
It is then immediately reheated at the same rate $R$ with a small 
magnetic field, $h=0.01$, which ensures linear response~\cite{fdt}. 
The zero field cooled magnetization, $M_{\rm zfc}(T)$, is then recorded. 
We then repeat the same protocol but this time include a stop
of duration $t_{\rm stop}$ at temperature $T_{\rm stop}$
during cooling. Upon reheating we now measure $M_{\rm stop}(T)$. 
The difference $ M_{\rm stop} - M_{\rm zfc} $ 
for  $t_{\rm stop} = 5\times 10^4$ and two different 
temperatures, $T_{\rm stop}=0.14$ and $T_{\rm stop} = 0.08$ 
is presented in Fig.~\ref{stop}. Curves have the typical shape found in 
experiments~\cite{matthieu} where the difference is maximal 
close to $T_{\rm stop}$ and becomes 0 at smaller and larger temperatures.
However the dip found here is broader than in experiments.
The fact that the curve has a bump at $T_{\rm stop}$ is akin to a memory 
effect since upon reheating the system remembers it has aged there. 
The fact that the difference diminishes at $T < T_{\rm stop}$
is a rejuvenation effect, because despite having aged 
close to $T_{\rm stop}$ the magnetization gets close to 
the reference curve. A full rejuvenation would imply that the difference 
really goes to 0. 
As above, we do find a good amount of memory but only a 
partial rejuvenation, because the  
separation of timescales is smaller than in experiments.

Recently, experiments have been carried out
on superspin glasses~\cite{petra}, in which the
microscopic timescale $t_0$ is considerably larger than in 
standard spin glasses ($t_0 \simeq 10^{-5} {\rm s}$), and so the 
timescales probed
are closer to those of our simulations than in experiments on standard spin
glasses. 
Interestingly, results of these experiments are quite similar to
those of our simulations and, in particular,
rejuvenation is only partial even when
temperature steps are quite substantial.
In our view, the claim made in 
Ref.~[\onlinecite{petra}] that rejuvenation is ``absent'' is merely 
a matter of definition, and we prefer to say 
that rejuvenation exists but is not full. Lack of full rejuvenation observed
in superspin glasses is consistent with our discussion
if the spread of timescales is not large enough for Eq.~(\ref{frcond}) or
(\ref{T1T2}) to be satisfied.  Implicit in Ref.~[\onlinecite{petra}] is the
idea that full rejuvenation necessarily implies strong temperature chaos.
However, we have argued here that full rejuvenation does not require
temperature chaos; just a large separation of time scales and a temperature
shift which is not too small.

\section{Summary}
\label{summary}

We have numerically studied memory and rejuvenation effects in the three
dimensional Heisenberg Edwards-Anderson 
spin glass.  The main difference between
experiments and simulations is that the simulations have a much less
pronounced separation of timescales.  As a result, memory effects in
temperature cycles can only be observed numerically for larger temperature
shifts than in experiments.

Similarly, we find that the timescale gap between experiments and simulations
also produces a qualitative difference when Ising and Heisenberg models are
compared.  In our numerical time window, the effect of a temperature shift on
aging dynamics is stronger for Ising systems than for Heisenberg systems, see
Fig.~\ref{teff} and Sec.~\ref{anisotropy}, since the dynamic correlation
length varies more strongly with $T$ in the Ising case.
However, in experiments, the
effect is the opposite~\cite{bert,india}. The numerical data shift, but only
slowly, towards experimental observations when larger timescales are
simulated, see Fig.~\ref{tefft}.

In addition to memory effects, we also observe partial rejuvenation for large
temperature shifts.  However, we find no sign of temperature chaos even for
much larger temperature shifts than in experiment.   The main difference
between the results of simulations, including ours, and experiments is that
simulations never find full rejuvenation. We have given an explanation of
this which does not involve temperature chaos but rather
depends on (i) the finite
cooling rate in experiments and (ii) the much larger
separation of time scales in experiments than in simulations.
The finite cooling rate also appears to be the explanation
for the `sub-aging' behavior found in experiments.

\begin{acknowledgments}
We thank J.-P. Bouchaud, F. Ricci-Tersenghi, E. Vincent 
and H. Yoshino for discussions. 
The work of APY is supported by the NSF through
grant DMR 0337049.
\end{acknowledgments}


\begin{thebibliography}{99}

\bibitem{I} L. Berthier and A.P. Young, 
Phys. Rev. B {\bf 69}, 184423 (2004).

\bibitem{review1} E. Vincent, J. Hamman, M. Ocio, J.-P. Bouchaud,
and L.F. Cugliandolo,
in {\it Complex behavior of glassy systems}, Ed.: M. Rubi (Springer Verlag,
Berlin, 1997).

\bibitem{review1_1}
P. Nordblad and P. Svendlidh, in
{\it Spin glasses and random fields}, Ed.: A.P. Young
(World Scientific, Singapore, 1998).

\bibitem{review3} L. Berthier, V. Viasnoff, O. White, V. Orlyanchik, 
and F. Krzakala, in {\it Slow relaxation and non equilibrium dynamics 
in condensed matter},  Eds: J.-L. Barrat, J. Dalibard, M. Feigelman, 
J. Kurchan (Springer, Berlin, 2003). 


\bibitem{dupuis} V. Dupuis, E. Vincent, J.-P. Bouchaud, J. Hammann,
A. Ito, and H.A. Katori, Phys. Rev. B {\bf 64}, 174204 (2001).

\bibitem{yosh4}  P.E. J\"onsson, H. Yoshino, and
P. Nordblad, Phys. Rev. Lett. {\bf 89}, 097201 (2002).

\bibitem{yosh3}
P.E. J\"onsson, R. Mathieu, P. Nordblad, H. Yoshino, H. Aruga Katori, 
and A. Ito, Phys. Rev. B {\bf 70}, 174402 (2004).

 \bibitem{bert} F. Bert, V. Dupuis, E. Vincent, J. Hammann, and
J.-P. Bouchaud,  Phys. Rev. Lett. {\bf 92}, 167203 (2004).

\bibitem{cycle} P. R\'efr\'egier, E. Vincent, J. Hammann, 
and M. Ocio, J. Phys. (France) {\bf 48}, 1533 (1987).

\bibitem{physica}
J. Hammann, M. Lederman, M. Ocio, R. Orbach, and E. Vincent,
Physica A {\bf 185}, 278 (1992). 

\bibitem{dip}
K. Jonason, E. Vincent, J. Hammann, J.-P. Bouchaud, and 
P. Nordblad, Phys. Rev. Lett. {\bf 81}, 3243 (1998).

\bibitem{petra} P.E. J\"onsson, H. Yoshino, H. Mamiya, and H. Takayama, 
cond-mat/0405276.

\bibitem{kovacs}
M. Sasaki, V. Dupuis, J.-P. Bouchaud, and E. Vincent,
Eur. Phys. J. B {\bf 29}, 469 (2002).

\bibitem{india} V. Dupuis, F. Bert, J.-P. Bouchaud, J. Hammann, F. Ladieu,
D. Parker, and E. Vincent, cond-mat/0406721. 

\bibitem{matthieu} R. Matthieu, P.E. J\"onsson, D.N.H. Nam, 
and P. Nordblad, Phys. Rev. B {\bf 63}, 092401 (2001). 


\bibitem{struik} L.C.E. Struik, {\it Physical aging
in amorphous polymers and other materials} (Elsevier, Houston, 1978).

\bibitem{frustre}
A.S. Wills, V. Dupuis, E. Vincent, J. Hammann, and R. Calemczuk,
Phys. Rev. B {\bf 62}, 9264 (2000).

\bibitem{leheny1}
R.L. Leheny and S.R. Nagel, 
Phys. Rev. B {\bf 57}, 5154 (1998).

\bibitem{leheny2} 
H. Yardimci and R.L. Leheny, Europhys. Lett. {\bf 62}, 203 (2003).

\bibitem{beta}
A.V. Kityk, M.C. Rheinst\"adter, K. Knorr, and H. Rieger,
Phys. Rev. B {\bf 65}, 144415 (2002).

\bibitem{KLT}
F. Alberici-Kious, J.-P. Bouchaud, L.F. Cugliandolo, P. Doussineau,
and A. Levelut, Phys. Rev. Lett. {\bf 81}, 4987 (1998).

\bibitem{bellon}
L. Bellon, S. Ciliberto, C. Laroche, 
Eur. Phys. J. B {\bf 25}, 223 (2002).

\bibitem{ferro}
E.V. Colla, L.K. Chao, and M.B. Weissman, Phys. Rev. B {\bf 63}, 
134107 (2001).

\bibitem{ferro2}
O. Kircher and R. B\"ohmer, Eur. Phys. J. B {\bf 26}, 329 (2002).

\bibitem{electron}
A. Vaknin, Z. Ovadyahu, and M. Pollak, Phys. Rev. B {\bf 65}, 
134208 (2002).


\bibitem{heiko}
H. Rieger, J. Phys. I (France) {\bf 4}, 883 (1994). 

\bibitem{rieger} J. Kisker, L. Santen, M. Schreckenberg, and H. Rieger, 
Phys. Rev. B {\bf 53}, 6418 (1996).

\bibitem{yosh2} T. Komori, H. Yoshino, 
and H. Takayama, J. Phys. Soc. Jpn {\bf 69}, 1192 (2000).

\bibitem{BB1}
L. Berthier and J.-P. Bouchaud, Phys. Rev. B {\bf 66}, 054404 (2002).

\bibitem{BB2}
L. Berthier and J.-P. Bouchaud,
Phys. Rev. Lett. {\bf 90}, 059701 (2003).

\bibitem{ricci}  M. Picco, F. Ricci-Tersenghi, F. Ritort,
Phys. Rev. B {\bf 63}, 174412 (2001).

\bibitem{ricci2} A. Maiorano, E. Marinari, and 
F. Ricci-Tersenghi, preprint cond-mat/0409577.

\bibitem{jimenez} S. Jimenez, V. Martin-Mayor and S. Perez-Gaviro, 
preprint cond-mat/0406345.


\bibitem{fh} D.S. Fisher and D. A. Huse, 
Phys. Rev. Lett. {\bf 56}, 1601 (1986);
Phys. Rev. B {\bf 38}, 
373 (1988); {\it ibid.} {\bf 38}, 386 (1988).

\bibitem{jorge}
L.F. Cugliandolo and J. Kurchan,
Phys. Rev. B {\bf 60}, 922 (1999).

\bibitem{jp} J.-P. Bouchaud, in
{\it Soft and fragile matter: 
non equilibrium dynamics, metastability and flow}, Eds.: M.E. Cates and
M.R. Evans (Institute of Physics Publishing, Bristol, 2000).

\bibitem{encorejp} J.-P. Bouchaud, V. Dupuis, J. Hammann, and E. Vincent, 
Phys. Rev. B {\bf 65}, 024439 (2001).

\bibitem{yosh}
H. Yoshino, A. Lema\^{\i}tre, and J.-P. Bouchaud,
Eur. Phys. J. B {\bf 20}, 367 (2001).

\bibitem{surf} L. Berthier and P.C.W. Holdsworth,
Europhys. Lett. {\bf 58}, 35 (2002).

\bibitem{yosh5_1} F. Scheffler, H. Yoshino, and P. Maas, 
Phys. Rev. B {\bf 68}, 060404 (2003).

\bibitem{martin} M. Sasaki and O. C. Martin, 
Phys. Rev. Lett. {\bf 91}, 097201 (2003).

\bibitem{yosh5} H. Yoshino, J. Phys. A {\bf 36}, 10819 (2003).


\bibitem{kawa} H. Kawamura, Phys. Rev. Lett. {\bf 80}, 5421 (1998);
Phys. Rev. Lett. {\bf 90}, 237201 (2003).

\bibitem{II} L. Berthier and A.P. Young,
J. Phys.: Condens. Matter {\bf 16}, S729 (2004).


\bibitem{dlogS}
H. Takayama and K. Hukushima,
J. Phys. Soc. Jpn. {\bf 73}, 2077 (2004).

\bibitem{cool} V.S. Zotev, G.F. Rodriguez, G.G. Kenning, R. Orbach, 
E. Vincent, and J. Hammann, Phys. Rev. B {\bf 67}, 184422 (2003).


\bibitem{bm1} 
A.J. Bray and M.A. Moore, Phys. Rev. Lett. {\bf 58}, 57 (1987).

\bibitem{timo}
T. Aspelmeier, A.J. Bray, and M.A. Moore, Phys. Rev. Lett. {\bf 89},
197202 (2002).

\bibitem{hajime} 
M. Sasaki, K. Hukushima, H. Yoshino, H. Takayama, preprint
cond-mat/0411138.

\bibitem{kr}
F. Krzakala, Europhys. Lett. {\bf 66}, 847 (2004).

\bibitem{mauro} L. Berthier, P.C.W. Holdsworth and M. Sellitto,
J. Phys. A {\bf 34}, 1805 (2001).

\bibitem{www}
See:
http://www.sheilaomalley.com/archives/002514.html

\bibitem{cooling} H. Yoshino and P.E. J\"onsson
preprint cond-mat/0407459.

\bibitem{fdt} L. Berthier and A.P. Young (unpublished).

\end{thebibliography}
\end{document}